%% file: main.tex
\documentclass[11pt]{article}

\PassOptionsToPackage{table}{xcolor}
\usepackage[preprint]{acl}

\usepackage{times}
\usepackage{latexsym}

\usepackage[T1]{fontenc}

\usepackage[utf8]{inputenc}

\usepackage{microtype}

\usepackage{inconsolata}
\usepackage[normalem]{ulem}
\usepackage{alltt}

\usepackage{graphicx}
\usepackage{bbding}
\usepackage{amssymb}
\usepackage{multirow}
\usepackage{booktabs}
\usepackage{tcolorbox}
\tcbuselibrary{skins, breakable, listings}

\definecolor{promptbg}{RGB}{252, 252, 252}
\definecolor{promptframe}{RGB}{220, 220, 220}
\definecolor{prompttitle}{RGB}{80, 80, 80}

\newtcblisting{promptbox}[1][]{%
  enhanced,
  breakable,
  colback=promptbg,
  colframe=promptframe,
  arc=3pt,
  boxrule=0.8pt,
  left=10pt,
  right=10pt,
  top=10pt,
  bottom=10pt,
  before skip=15pt,
  after skip=15pt,
  listing only,
  listing options={
    basicstyle=\ttfamily\small,
    columns=fullflexible,
    breaklines,
    breakindent=0pt,
    breakatwhitespace=false,
    aboveskip=0pt,
    belowskip=0pt,
    showstringspaces=false,
  },
  \if\relax\detokenize{#1}\relax\else
    title={#1},
    coltitle=prompttitle,
    fonttitle=\bfseries\small,
    attach boxed title to top left={xshift=4mm, yshift=-3mm},
    boxed title style={
      colback=white,
      colframe=promptframe,
      boxrule=0.8pt,
      arc=2pt,
    },
    top=12pt,
  \fi
}

\definecolor{casebg}{RGB}{246, 251, 247}
\definecolor{caseframe}{RGB}{43, 112, 78}
\definecolor{casetitlebg}{RGB}{43, 112, 78}
\definecolor{casetitlefg}{RGB}{255, 255, 255}

\newtcolorbox{casebox}[1][Case]{%
  enhanced,
  breakable,
  colback=casebg,
  colframe=caseframe,
  coltitle=casetitlefg,
  colbacktitle=casetitlebg,
  fonttitle=\bfseries,
  boxrule=0.9pt,
  arc=1.6mm,
  left=8pt,
  right=8pt,
  top=7pt,
  bottom=7pt,
  before skip=10pt,
  after skip=10pt,
  width=\linewidth,
  title={#1}
}

\definecolor{takeawaybg}{RGB}{237, 242, 247}
\definecolor{takeawayframe}{RGB}{154, 170, 184}
\newcounter{takeaway}
\newenvironment{takeawaybox}{%
  \vspace{0.35em}%
  \refstepcounter{takeaway}%
  \begin{tcolorbox}[
    enhanced,
    colback=takeawaybg,
    colframe=takeawayframe,
    boxrule=0.5pt,
    arc=2pt,
    left=8pt,
    right=8pt,
    top=6pt,
    bottom=6pt,
    before skip=0pt,
    after skip=0pt,
  ]
  \textbf{Takeaway~\thetakeaway.}\quad\ignorespaces
}{%
  \end{tcolorbox}%
  \vspace{0.35em}%
}

\hypersetup{
  colorlinks=true,
  linkcolor=blue!65!black,
  citecolor=blue!65!black,
  urlcolor=blue!65!black
}

\title{Are Diffusion Language Models Good Database Analysts?}

\author{%
  Peixian~Ma\textsuperscript{1,}\textsuperscript{2},
  ~Xialie~Zhuang\textsuperscript{3},
  ~Jiantao~Tan\textsuperscript{4},
  ~Changlun~Li\textsuperscript{1,}\textsuperscript{2},
  ~Ruirui~Chen\textsuperscript{5},
  ~Chengwei~Qin\textsuperscript{1} 
  \\[4pt]
  \textsuperscript{1}HKUST(GZ) \quad
  \textsuperscript{2}Paradoox AI \quad
  \textsuperscript{3}UCAS \quad
  \textsuperscript{4}SYSU \quad
  \textsuperscript{5}A*STAR 
  \\[4pt]
  \texttt{pma929@connect.hkust-gz.edu.cn}
}

\begin{document}

\maketitle

\begin{abstract}
  Diffusion Language Models~(DLMs) enable iterative global refinement, but their effectiveness in Natural Language to SQL~(NL2SQL) remains unclear because heterogeneous architectures and evaluation setups hinder fair comparison with autoregressive~(AR) models. Their iterative denoising process permits sequence-wide revision, offering a different way to handle the long-range structural dependencies in SQL. We present a systematic study of DLM-based NL2SQL under a unified framework that standardizes generation and execution across DLM architectures. We further introduce \texttt{SQL-D1}, an agentic inference framework combining database-aware retrieval, multi-trajectory denoising, execution-guided verification, and selection based on execution denotations. Across diverse benchmark settings, strong DLMs are competitive with some open-source AR models and offer flexible trade-offs between efficiency and accuracy, although they still trail the strongest AR systems on difficult queries. Test-time scaling yields substantial gains, whereas standard supervised fine-tuning yields unstable outcomes across the evaluated DLMs. Stronger model variants exhibit fewer schema-grounding errors, but failures involving complex clauses and query nesting persist. These results clarify the capabilities and limitations of DLMs for database analysis and provide a standardized foundation for future work. Our code is available at \url{https://anonymous.4open.science/r/SQL-D1-82DF}.
\end{abstract}

\section{Introduction}
\label{sec:introduction}

\begin{figure}[t!]
    \centering
    \includegraphics[width=\linewidth]{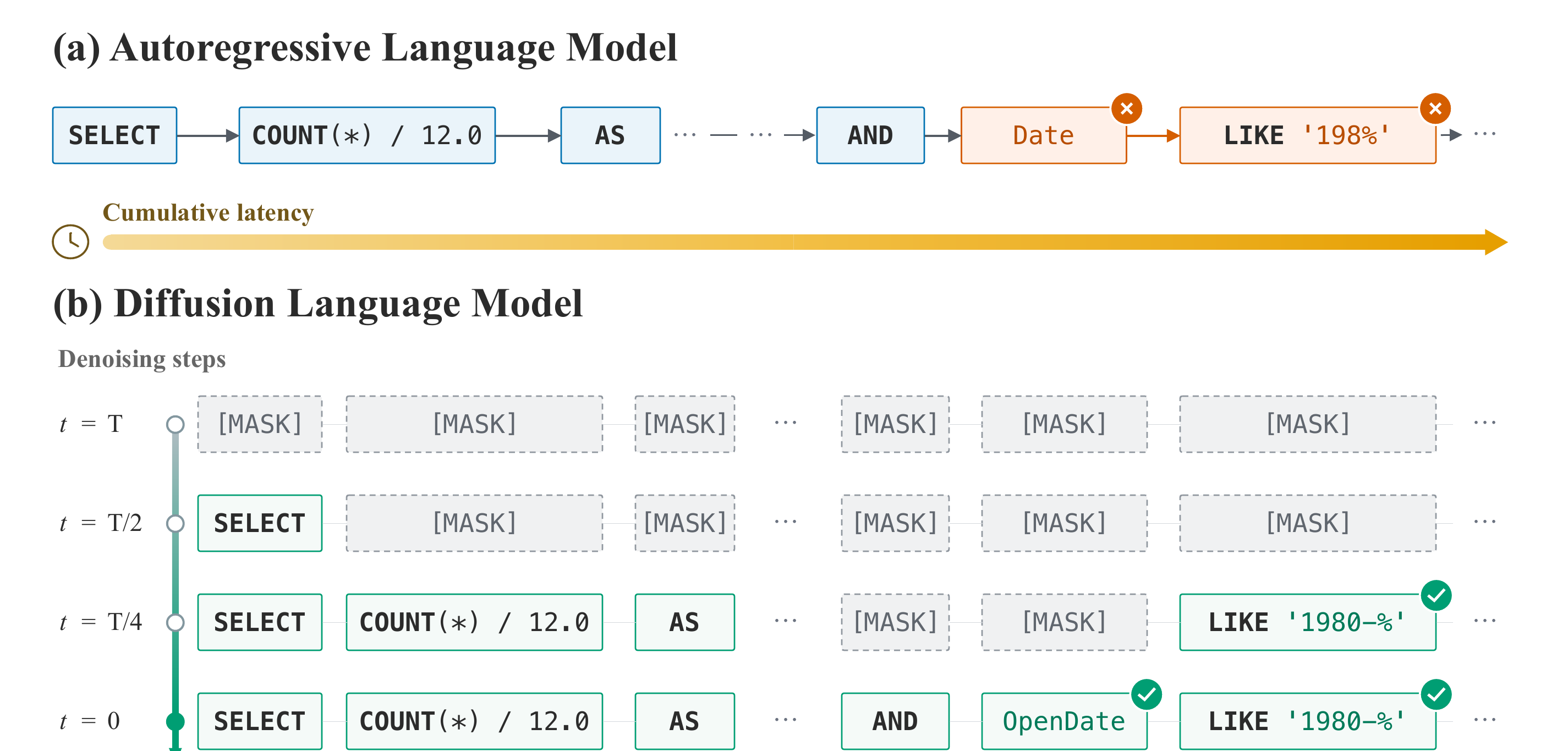}
    \caption{Demonstration of our work. While AR models decode left-to-right and risk sequential error propagation, DLMs perform iterative global denoising. Whether this non-autoregressive refinement translates into a net advantage in database analysis is the core empirical question investigated in this work.}
    \label{fig:motivation_intro}
\end{figure}

Natural Language to SQL~(NL2SQL) serves as a crucial bridge for evaluating the database analysis capabilities of language models~\citep{DBLP:journals/corr/abs-2408-05109, hong2024next}. While recent progress driven by large language models~(LLMs) has significantly enhanced generation accuracy and schema comprehension~\citep{dailsql, omnisql, chasesql, sqlr1}, these systems primarily rely on the autoregressive~(AR) paradigm. Given that SQL is a highly structured and non-linear language, the rigid left-to-right decoding process of AR models makes them particularly susceptible to sequential error propagation. Errors introduced early in the generation process often accumulate throughout the sequence. The lack of mechanisms for global structural revision ultimately limits the effectiveness of AR models in performing complex database analysis tasks.

Recently, diffusion language models~(DLMs) have emerged as a promising alternative to unidirectional decoding~\citep{dlmsurvey, arriola2025interpolating, llada}. As shown in Figure~\ref{fig:motivation_intro}, DLMs replace unidirectional decoding with an iterative denoising process, enabling global refinement of sequence representations through multiple parallelizable steps. This mechanism \emph{may} help NL2SQL by allowing models to revise semantic inconsistencies and grounding errors across the query, but whether the potential translates into a net advantage over AR models is an empirical question. Leading DLMs~\citep{llada, wedlm, dream, llada20, llada21} have demonstrated performance comparable to LLMs in mathematics and code generation. Consequently, a systematic investigation into whether these models can effectively address structural challenges in database reasoning represents a significant research direction.

Despite this considerable potential, the application of DLMs to the NL2SQL domain is currently hindered by the absence of a standardized evaluation infrastructure. The DLM ecosystem remains highly fragmented, encompassing divergent methodologies such as discrete diffusion, continuous diffusion, and mask reconstruction. These approaches are associated with different training objectives, decoding procedures, and sampling strategies. This heterogeneity creates a substantial methodological gap, preventing equitable and quantitative comparisons among various diffusion architectures and traditional AR baselines. It also impedes the direct adaptation of established AR-centric enhancement techniques to the diffusion paradigm. Therefore, without unified evaluation protocols and a specialized integration framework, the true capability boundaries and practical viability of DLMs for structured database analysis remain unclear.

To address these challenges and research gaps, we aim to answer the following research questions: 
\begin{itemize}
    \item \textbf{\textit{Q1:}} How do DLMs perform relative to AR models in NL2SQL tasks?
    \item \textbf{\textit{Q2:}} How can we improve performance of DLMs in NL2SQL tasks?
    \item \textbf{\textit{Q3:}} What are the distinctive failure modes and scaling behaviors of DLMs in structured generation?
\end{itemize}

\noindent Takeaways. In summary, our work makes the following contributions and provides takeaways:

(i)~DLMs demonstrate highly competitive performance and structural robustness in database analysis tasks. Advanced DLMs, such as LLaDA2.1-mini, achieve state-of-the-art execution accuracy among non-autoregressive models, surpassing comparable autoregressive baselines.

(ii)~DLMs demonstrate significant performance improvements through test-time scaling and provide a flexible balance between computational efficiency and accuracy.

(iii)~Agentic coordination effectively unleashes the latent potential of DLMs, driving substantial accuracy gains across primary benchmarks.

(iv)~Supervised fine-tuning methods exhibit instability when applied to DLMs on NL2SQL tasks, leading to inconsistent performance gains.

(v)~Scaling up model size substantially reduces grounding errors in conditions and tables. However, structural errors in complex clauses and query nesting persist across various diffusion architectures.

\paragraph{Contributions.} In summary, our work makes the following contributions and provides takeaways:

(i) \textbf{Unified NL2SQL evaluation framework across diverse DLMs}: We establish a unified evaluation framework for DLM-based NL2SQL, defining rigorous protocols for cross-benchmark comparison. This standardization enables a systematic, quantitative assessment of DLM capability boundaries in structured database analysis under consistent generation and execution environments.

(ii) \textbf{Agentic DLM-based NL2SQL system}: We introduce \texttt{SQL-D1}, a novel synthesis framework that integrates database-aware context engineering, test-time scaling, and agentic optimization to enable more robust and effective NL2SQL reasoning with DLMs.

(iii) \textbf{Empirical analysis and takeaways of DLM-based NL2SQL}: We provide critical insights into the test-time scaling potential, post-training instability, and primary failure modes of DLMs, formalizing structured takeaways that characterize the efficiency-accuracy trade-offs and structural robustness of diffusion architectures.

\section{Methodology}
\label{sec:methodology}

\begin{figure*}[t!]
  \centering
  \includegraphics[width=\linewidth]{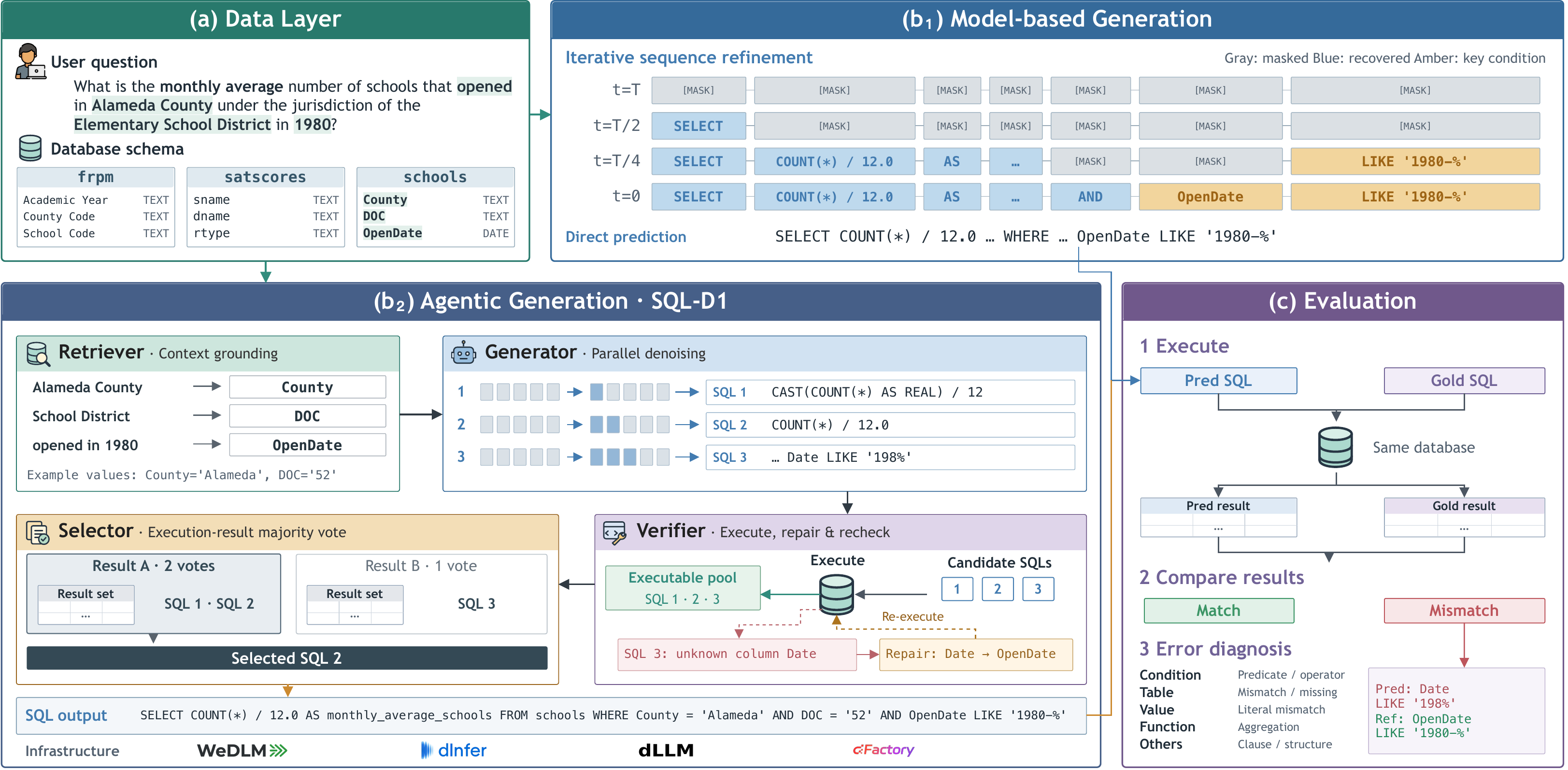}
  \caption{Overview of the proposed DLM-based NL2SQL framework. The framework comprises (a) a data layer, (b$_1$) direct model-based generation, (b$_2$) \texttt{SQL-D1} agentic generation with retrieval, parallel denoising, execution-guided verification, and selection based on execution denotations, and (c) execution-based evaluation with fine-grained error diagnosis.}
  \label{fig:framework}
\end{figure*}

\subsection{Overview}
\label{sec:method-overview}

As illustrated in Figure~\ref{fig:framework}, the unified evaluation framework standardizes interactions among users, databases, and heterogeneous DLM architectures through three integrated modules. The system employs a dual-track inference pipeline supporting both direct model-based generation and a structured agentic workflow comprising retrieval, generation, verification, and selection. The process concludes with a comprehensive evaluation engine that performs functional verification via SQL execution and provides fine-grained diagnostic analysis of error typologies.

\subsection{\texttt{SQL-D1}: Agentic NL2SQL with DLMs}
\label{sec:method-sql-d1}

Building on agentic NL2SQL systems for AR models~\citep{chasesql,macsql,agenterscalesql}, \texttt{SQL-D1} organizes DLM-based SQL generation into four stages: retrieval, generation, verification, and selection. The retriever provides database-specific context, and the generator produces candidate queries through multiple independent denoising trajectories. The verifier uses execution feedback to identify and, when possible, repair queries that fail to execute. The selector then groups executable candidates by execution result and selects a final query. Table~\ref{tab:agent_io} summarizes the inputs and outputs of the four agents.

\begin{table}[t]
\centering
\small
\setlength{\tabcolsep}{2.5pt}
\begin{tabular}{@{}l p{2.5cm} p{2.8cm} c@{}}
\toprule
\textbf{Agent} & \textbf{Input} & \textbf{Output} & \textbf{Exec.} \\
\midrule
$A_r$ & Question and current-database schema & Grounded context with one example value per field & No \\
$A_g$ & Grounded context and question & $k$ candidate SQLs from multi-trajectory denoising & No \\
$A_v$ & Candidates and execution feedback & Repaired or filtered executable candidates & Yes \\
$A_s$ & Executable candidates and denotations & One SQL~(majority denotation; shortest-SQL tie-break) & Yes \\
\bottomrule
\end{tabular}
\caption{Input/output of \texttt{SQL-D1} agents. Exec.\ indicates whether the agent runs SQL against the database.}
\label{tab:agent_io}
\end{table}

\input{Tables/main_eval.tex}

\paragraph{Retriever $A_r$.} The Retriever constructs database-specific context using the schema of the database associated with each question. It augments the schema metadata with one representative value per field, selected offline from that database. The same context construction procedure is applied to the DLM and AR models in the main comparison.

\paragraph{Generator $A_g$.} Conditioned on the retrieved context, the Generator performs test-time scaling by launching multiple concurrent denoising trajectories rather than a single left-to-right decode. This design uses the parallelizable refinement of DLMs and the high variance of single-pass diffusion generation. The resulting candidate pool is passed to verification and selection under a controllable inference budget.

\paragraph{Verifier $A_v$.} The Verifier introduces an agentic correction loop designed to ensure the formal validity and semantic alignment of the generated candidates. It leverages a feedback-driven mechanism to iteratively diagnose and rectify potential structural inconsistencies or execution failures. By transforming raw candidates into executable queries through targeted refinement, the Verifier bridges the gap between probabilistic generation and the deterministic requirements of SQL execution.

\paragraph{Selector $A_s$.} The Selector aggregates executable candidates by execution denotation rather than by exact SQL string match. The most frequent denotation group is selected; ties among equal-frequency groups are broken by choosing the shortest SQL. Non-executable candidates do not vote. This rule is consistent with EX as the primary metric and reduces variance across denoising trajectories.

\section{Experiments}
\label{sec:experiments}

\subsection{Experimental Setup}
\label{sec:experiments-datasets_and_metrics}

\paragraph{Benchmark Settings.} We conduct our experiments on six main NL2SQL benchmarks, including Spider~\citep{spider}, BIRD~\citep{bird}, Spider-DK~\citep{spider-dk}, Spider-Syn~\citep{spider-syn}, and Spider-Realistic~\citep{spider-realistic}. Detailed descriptions of these benchmarks are provided in Appendix~\ref{sec:appendix-detailed_evaluation_settings}.

\paragraph{Metric \& Baselines.} To ensure fair comparisons, we adhere to the standard evaluation metric established in prior works. Specifically, we use execution accuracy~(EX) as the primary metric across all benchmarks. We compare DLM families~(Dream, DiffuCoder, WeDLM, LLaDA) against several AR model baselines on the main NL2SQL benchmarks. Detailed models of these baselines are provided in Table~\ref{tab:main_results}, Appendix~\ref{sec:appendix-dlm-details} and Appendix~\ref{sec:appendix-llm-baselines}.

\paragraph{Environment.} All experiments in this study are conducted on a server running Ubuntu 22.04. This server is equipped with an Intel(R) Xeon(R) Platinum 8358 CPU @ 2.60 GHz, 512 GB of system memory, and 8 A100-80GB GPUs.

\subsection{Main Results}
\label{sec:experiments-main_results}

\paragraph{\textit{Q1: How do DLMs perform relative to AR models in NL2SQL tasks?}} We first compare DLMs and AR models on the main NL2SQL benchmarks.

\paragraph{Overall Performance Comparison.}~Table~\ref{tab:main_results} summarizes performance across diverse benchmarks. A primary observation is the rapid ascent of DLM capabilities, particularly within the LLaDA2 and WeDLM families. LLaDA2.1-mini attains the best EX among the evaluated DLMs, reaching 78.6\% Maj@8 on Spider-Test. A gap remains relative to high-end AR models such as GPT-4-Turbo~(84.2\%). Notably, under the same zero-shot prompting configuration, LLaDA2.1-mini~(58.7\%) outperforms several representative open-source AR models on the BIRD benchmark, such as Qwen2.5-Coder-7B~(58.2\%), Qwen2.5-7B~(56.4\%), and Qwen3-8B-Thinking~(51.8\%).

The robustness of DLMs is further demonstrated by their performance on schema-perturbed and domain-specific variants. On Spider-DK and Spider-Realistic, LLaDA2.1-mini achieves competitive accuracies of 67.5\% and 72.2\%, respectively.
These benchmarks introduce domain and schema variations, making them effective tests of whether a model relies on superficial lexical matching. LLaDA2.1-mini remains competitive on these variants, demonstrating robust performance under domain and schema variations.

\input{Tables/difficulty_eval.tex}

\begin{figure}[t!]
  \centering
  \includegraphics[width=\linewidth]{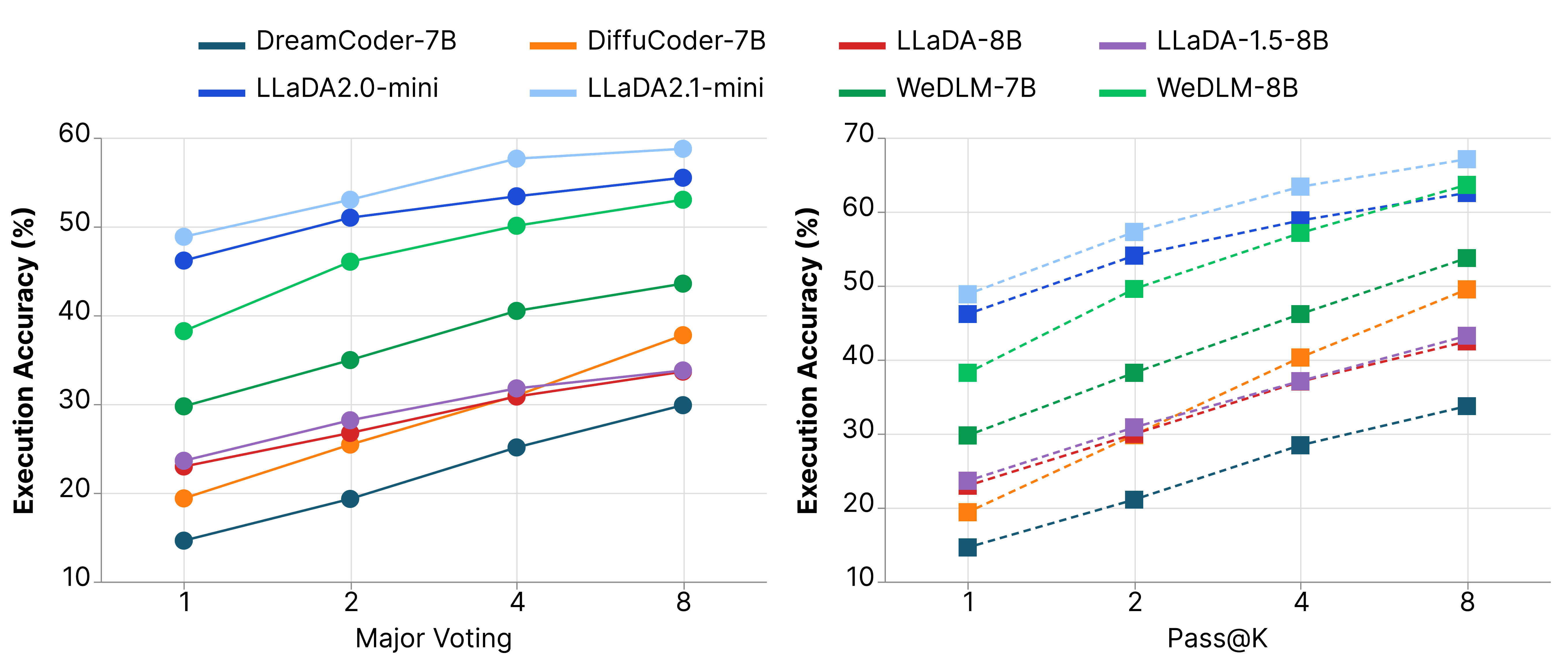}
  \caption{Performance comparison of DLMs with different test-time scaling strategies~(\textbf{left}: majority voting, \textbf{right}: pass@k) on BIRD-Dev dataset.}
  \label{fig:passk}
\end{figure}

\paragraph{Complexity Analysis.}~Table~\ref{tab:difficulty_eval} provides a fine-grained breakdown of execution accuracy by complexity level on BIRD-Dev. LLaDA2.1-mini demonstrates remarkable consistency, achieving 49.0\% on challenging queries, surpassing specialized NL2SQL methods such as CodeS-15B~(42.4\%) and DAIL-SQL~(43.1\%). Despite these gains, a performance ceiling remains when compared to top-tier models such as Qwen2.5-Coder-14B~(88.0\% on Spider-Test). This gap is most pronounced in extremely complex reasoning tasks, indicating that while DLMs are formidable contenders, further scaling and structural optimizations are required to reach the leading performance of autoregressive LLMs.

\begin{takeawaybox}
  Strong DLMs are competitive with some AR models, while still trailing the strongest AR systems on difficult queries.
\end{takeawaybox}

\begin{figure*}[t!]
  \centering
  \includegraphics[width=\linewidth]{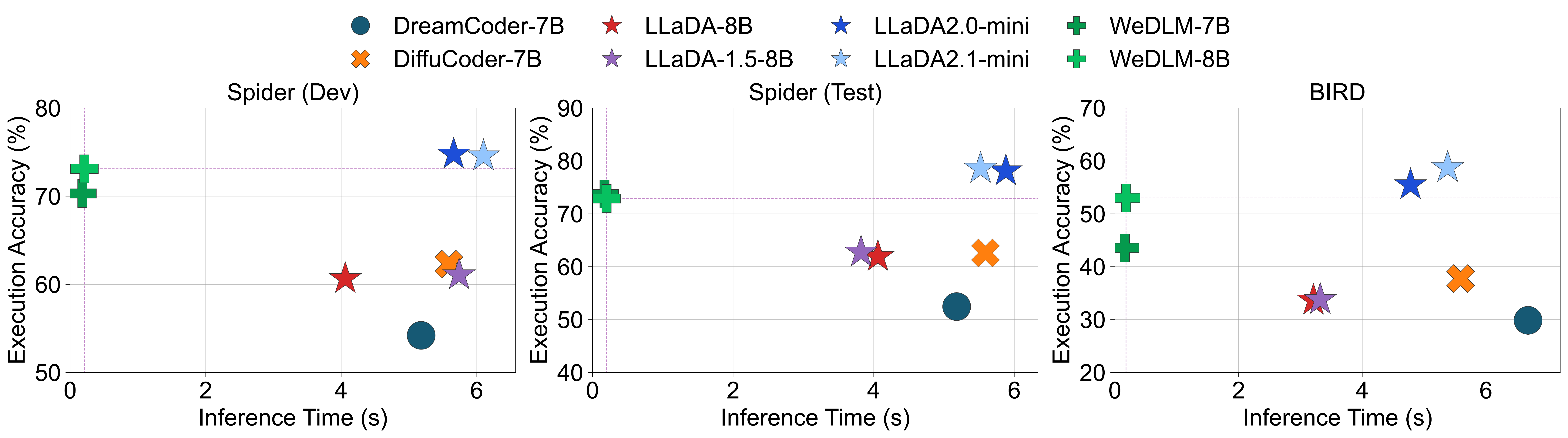}
  \caption{Efficiency and execution accuracy trade-offs of DLMs on Spider-Dev~(\textbf{left}), Spider-Test~(\textbf{middle}), and BIRD-Dev~(\textbf{right}) datasets.}
  \label{fig:efficiency_visualization}
\end{figure*}

\paragraph{\textit{Q2: How can we improve performance of DLMs in NL2SQL tasks?}} We further investigate the potential of DLMs in NL2SQL tasks by analyzing the impact of test-time scaling, the integration of an agentic framework, and the training strategies.

\input{Tables/ablation_sqld1_eval.tex}

\paragraph{Analysis of Test-time Scaling \& Efficiency.} To further investigate the potential of DLMs in NL2SQL tasks, we analyze the impact of test-time scaling on performance. As illustrated in Figure~\ref{fig:passk}, all DLMs exhibit a significant and sustained upward trend in performance as the number of sampled candidates $k$ increases. For instance, WeDLM-8B achieves a pass@8 score of 63.6\% on the BIRD-Dev set, representing an improvement of over 25 percentage points compared to its pass@1 accuracy of 38.2\%. With the majority voting strategy, the performance of LLaDA2.1-mini increases steadily from 48.8\% to 58.7\% at $k=8$. This robust scaling effect demonstrates that DLMs can effectively mitigate the stochasticity of single-pass generation by increasing the computational budget during inference. Furthermore, the substantial gap between pass@$k$ and Maj@$k$ highlights the considerable upper bound potential of DLMs. With the integration of more sophisticated re-ranking mechanisms, their performance is poised to approach, and perhaps even exceed, that of leading AR models.

Figure~\ref{fig:efficiency_visualization} illustrates the efficiency and execution accuracy trade-offs of various DLMs, where the WeDLM and LLaDA2 series clearly define the Pareto frontier for the DLM-based NL2SQL system. Specifically, the WeDLM series exhibits exceptional inference efficiency, achieving a single-pass latency of approximately 0.2 seconds per query, which is nearly 30 times faster than the LLaDA2 series. In contrast, the LLaDA2 series delivers the highest execution accuracy among all evaluated DLMs, reaching 78.6\% on Spider-Test and 58.7\% on BIRD-Dev, albeit at a higher computational cost per pass. This versatile trade-off between ultra-low latency and high-fidelity generation underscores the competitive advantage of DLMs across diverse real-time application scenarios. Consequently, our subsequent analyses concentrate on these two model families to further characterize their reasoning capabilities and failure modes. 

Additionally, we present a budget-matched DLM vs.\ AR test-time scaling comparison in Appendix~\ref{sec:appendix-matched_scaling}, a matched \texttt{SQL-D1} ablation on an AR backbone in Appendix~\ref{sec:appendix-ar_sqld1}, and a comparison of efficiency, execution accuracy, and diffusion rendering hyperparameters on BIRD-Dev in Appendices~\ref{sec:appendix-efficiency_vs_baselines} and~\ref{sec:appendix-diffusion_rendering_params}.

\begin{takeawaybox}
  DLMs demonstrate significant performance gains through test-time scaling and offer a versatile trade-off between efficiency and accuracy, with the WeDLM and LLaDA2 series establishing the Pareto frontier for diffusion-based NL2SQL reasoning.
\end{takeawaybox}

\paragraph{Analysis of \texttt{SQL-D1}.} Table~\ref{tab:ablation_enhancement_eval} presents the ablation results for the \texttt{SQL-D1} framework, evaluating the contributions of each agentic module across multiple benchmarks. The results indicate that the integration of database-aware retrieval, iterative verification, and robust selection strategies consistently yields performance gains, with the full configuration substantially outperforming the baselines. Specifically, the execution accuracy for WeDLM-7B and WeDLM-8B increases to 70.8\% and 75.0\% on Spider-Dev, respectively, and a similar upward trend is observed on the BIRD-Dev dataset. These results suggest that the structured coordination of specialized agents effectively reduces the stochasticity inherent in diffusion-based generation and improves the structural grounding necessary for precise SQL generation.

The integration of the Retriever ($A_r$) offers a consistent baseline improvement by grounding the diffusion process in accurate schema metadata and database literals. The most significant gains, however, emerge with the introduction of the Verifier ($A_v$) and Selector ($A_s$), which utilize test-time scaling and iterative refinement. The agentic correction loop in $A_v$ effectively addresses structural and execution failures, while the robust aggregation in $A_s$ stabilizes the output by selecting the most consistent query from the diverse candidate pool.

Notably, \texttt{SQL-D1} demonstrates robust performance on challenging variants such as Spider-DK and Spider-Realistic. For example, on Spider-DK, the framework boosts WeDLM-7B's performance from 42.5\% to 62.0\%, an improvement of nearly 20 percentage points. This result indicates that the structured agentic pipeline is especially effective at addressing domain-specific knowledge gaps and linguistic variations that often hinder monolithic models. The combination of grounded retrieval, diverse exploration, and feedback-driven refinement ultimately enables DLMs to reach new levels of precision in complex NL2SQL tasks.

\input{Tables/ablation_training_eval.tex}

\begin{takeawaybox}
  The \texttt{SQL-D1} framework demonstrates that agentic coordination, which combines grounded retrieval, test-time scaling, and iterative refinement, effectively unlocks the latent potential of DLMs in NL2SQL tasks and delivers significant performance improvements as well as enhanced robustness across diverse and challenging database scenarios.
\end{takeawaybox}

\paragraph{Analysis of Training Strategies.} Table \ref{tab:ablation_training_eval} presents the impact of post-training strategies on DLM performance. We conduct Supervised Fine-Tuning~(SFT) using the BIRD-Train-9K dataset on several representative models to assess their responsiveness to domain-specific tuning. The experimental results indicate that the current post-training paradigm for DLMs is notably unstable when compared with established protocols for autoregressive models. For example, WeDLM-7B-Instruct achieves only a marginal increase of 0.5\% in execution accuracy, while both WeDLM-8B-Instruct and LLaDA2.0-mini exhibit performance degradation, decreasing by 2.1\% and 1.8\%, respectively. These findings suggest that direct SFT on database-specific datasets may unintentionally disrupt the internal representations or the denoising trajectories acquired during the initial training process. The observed inconsistencies emphasize the need for further research into specialized post-training methodologies that are better aligned with the iterative generation mechanism of DLMs. The development of robust alignment and fine-tuning techniques remains a primary challenge for improving the domain-specific reasoning capabilities of diffusion-based language models in NL2SQL tasks.

\begin{takeawaybox}
  The instability and inconsistent performance outcomes of conventional supervised fine-tuning on DLMs underscore the critical need for specialized post-training methodologies that are specifically aligned with the iterative denoising mechanisms of diffusion language models.
\end{takeawaybox}

\begin{figure}[t!]
  \centering
  \includegraphics[width=0.93\linewidth]{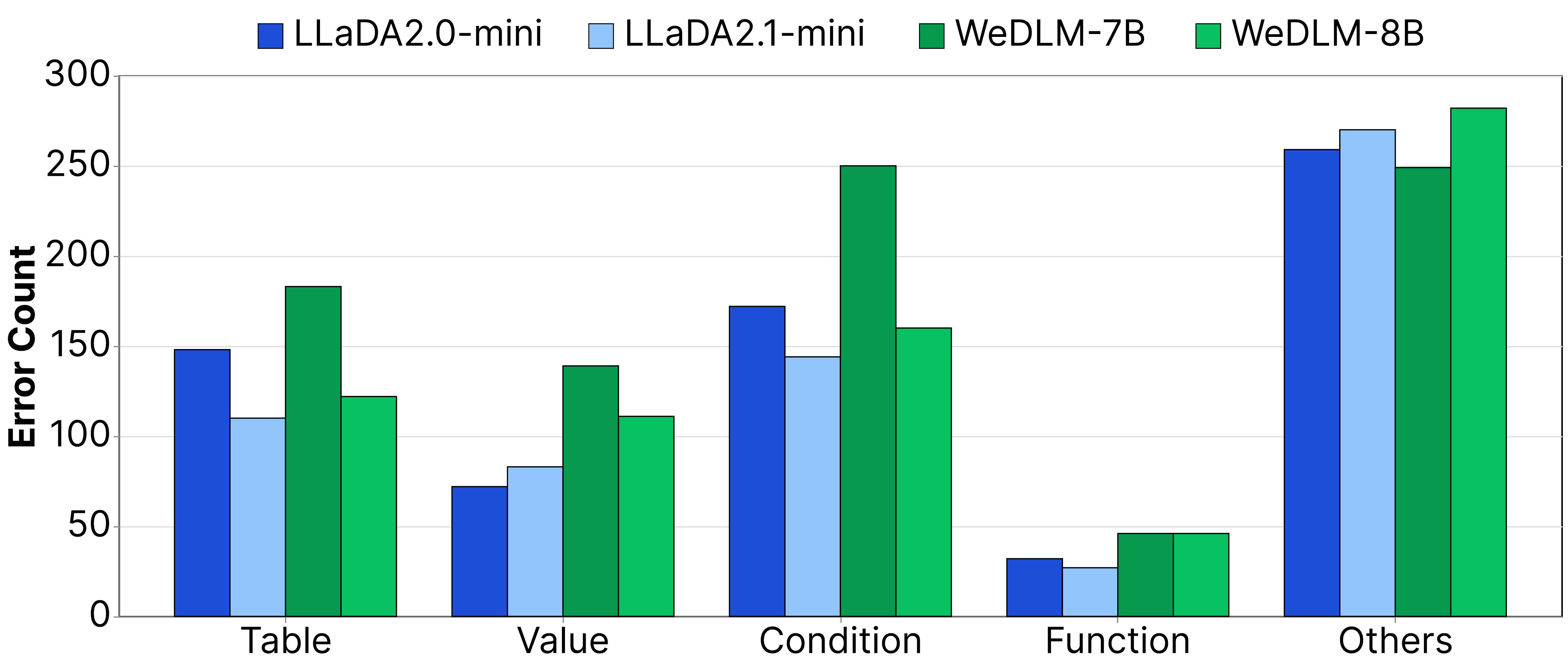}
  \caption{Error distribution of LLaDA2 and WeDLM series models on BIRD-Dev dataset.}
  \label{fig:error_distribution}
\end{figure}

\paragraph{\textit{Q3: What are the distinctive failure modes and scaling behaviors of DLMs in structured generation?}} We further investigate the failure modes and scaling behaviors of DLMs in structured generation by conducting a fine-grained error analysis.

\paragraph{Error Analysis.} To better understand the failure modes of DLMs in structured reasoning and identify specific areas for improvement, we conduct a fine-grained error analysis. We specifically compare the diagnostic behavior of the LLaDA2 and WeDLM series models to uncover how different diffusion architectures handle complex SQL generation. We categorize the errors of DLMs in NL2SQL tasks into five categories: \textbf{Table}~(e.g.,~\textit{table mismatch}, \textit{table missing}), \textbf{Value}~(e.g.,~\textit{value mismatch}), \textbf{Condition}~(e.g.,~\textit{attribute error}, \textit{operator error}), \textbf{Function}~(e.g.,~\textit{incorrect use of aggregate functions like SUM, AVG}), and \textbf{Others}~(e.g.,~\textit{clause missing}).

Figure~\ref{fig:error_distribution} demonstrates the distribution of errors across these categories. As illustrated, \textbf{Condition} and \textbf{Table} errors constitute the primary failure modes for both model families. This distribution highlights the challenges of precise schema linking and complex predicate formulation in a non-autoregressive setting. Specifically, WeDLM-7B exhibits the highest frequency of \textbf{Condition} (250) and \textbf{Table} (183) errors. These counts are significantly mitigated in WeDLM-8B (160 and 122, respectively), suggesting that increased model scale and refined training objectives improve structural grounding by enhancing the mapping of natural language intents to database entities. The LLaDA2 series demonstrates superior performance in handling \textbf{Value} mapping, where LLaDA2.0-mini incurs only 72 such errors compared to 139 for WeDLM-7B. Furthermore, the transition from LLaDA2.0 to LLaDA2.1-mini yields a notable reduction in \textbf{Table} errors (from 148 to 110), indicating an enhanced capability in discerning relevant database entities during the iterative denoising process. The prevalence of \textbf{Others} errors across all models is particularly telling because these failures predominantly involve missing structural clauses and incorrect subquery nesting. These specific failure modes provide direct evidence of the difficulty that DLMs face in maintaining the global structural coherence and long-range dependencies required for complex SQL generation. Additionally, we provide a detailed error taxonomy and case study in Appendix~\ref{sec:appendix-case_study}.

\begin{takeawaybox}
  Condition and Table errors constitute the primary failure modes for DLMs and are significantly mitigated by increased model scale; however, structural failures related to clauses and nesting do not exhibit corresponding improvements with increased model capacity.
\end{takeawaybox}

\section{Related Work}
\label{sec:related_work}

\paragraph{Diffusion Language Models.} DLMs advance language modeling by replacing autoregressive decoding with iterative sequence or block refinement, incorporating masked denoising and discrete diffusion formulations~\citep{arriola2025interpolating,dlmsurvey,yu2025discrete,mounier2025reviewremaskrefiner3,vonrütte2025generalizedinterpolatingdiscretediffusion, yang2026continuousdiffusionscalescompetitively, zhang2026expertchoiceroutingenablesadaptive}. Recent work further explores scaling DLMs through adaptation from AR models~\citep{gong2025scaling}. Leading models such as LLaDA~\citep{llada}, Dream~\citep{dream}, and WeDLM~\citep{wedlm} demonstrate that global denoising can match or surpass sequential generation in structured domains including mathematics and code~\citep{nie2025scalingmaskeddiffusionmodels,ni2025trainingoptimallargediffusion, zhang2025corrective}. Nevertheless, the DLM ecosystem remains fragmented due to divergent training objectives and sampling strategies, which hinders standardized evaluation across models~\citep{wedlm,dinfer,zhou2026dllm}. To address this issue, this work introduces a unified framework that standardizes generation and execution protocols, thereby enabling the first systematic characterization of capability boundaries and diagnostic behaviors for diffusion models in structured database analysis.

\paragraph{Natural Language to SQL.} NL2SQL serves as a primary benchmark for evaluating database reasoning capabilities~\citep{DBLP:journals/corr/abs-2408-05109, hong2024next}. The field has evolved from specialized encoders to large language models~\citep{c3sql,supersql,dtssql}, with a focus on generation accuracy~\citep{codes,omnisql,opensearchsql,sqlr1,route,Soni_2026,marssql}, schema comprehension~\citep{resdsql,xiyansql,maamari2024death,leafsql,diver}, and executable semantic alignment~\citep{dailsql,ma2024plug,memosql}. Recent developments in agentic pipelines have enabled integration of test-time scaling and tool utilization within NL2SQL workflows~\citep{chasesql,mctssql,sqlo1,agenterscalesql,cai2026text2sqlflow}. Existing agentic designs primarily target AR models. We study how retrieval, generation, verification, and selection interact with iterative denoising in DLM-based NL2SQL. Current evaluation protocols still primarily address autoregressive architectures~\citep{rajkumar2022evaluating}. This work therefore standardizes evaluation for DLMs, integrates \texttt{SQL-D1} as an inference-time system, and analyzes fine-grained error typologies.

\section{Conclusion}
\label{sec:conclusion}

In this work, we establish a unified framework for systematically evaluating DLMs in NL2SQL. Strong DLMs are competitive with some open-source AR models and offer flexible trade-offs between efficiency and accuracy, while still trailing the strongest AR systems on difficult queries. The \texttt{SQL-D1} framework further improves execution accuracy through database-aware retrieval and execution-grounded test-time scaling, with matched ablations showing similar full-pipeline gains on an AR backbone. Despite fewer schema-grounding errors in stronger model variants, persistent structural errors and unstable SFT outcomes highlight the need for optimization methods tailored to iterative denoising.

\clearpage

\section*{Limitations}
\paragraph{Dialect Constraints.} The current scope of our empirical evaluation is primarily restricted to the SQLite dialect, which serves as the foundational environment for the proposed unified protocol. Although SQLite is widely utilized for benchmarking structured reasoning tasks, we recognize that other database systems may present distinct syntactic and execution challenges. We are committed to an iterative development process that aims to broaden the compatibility of our framework across a more diverse range of database management systems in future versions.

\paragraph{Post-training.} Our post-training analysis evaluates standard SFT on BIRD-Train-9K and reveals unstable performance across the evaluated DLMs. Developing DLM-specific post-training methods remains an important direction for future work.

\paragraph{Model Coverage.} This work focuses on a specific selection of representative DLM architectures. As the field of DLM is undergoing rapid advancement, certain newly released models have not yet been included in our current analysis. We view this study as an evolving effort and intend to incorporate the latest developments in the DLM ecosystem through continuous iterations of our suite to provide a more comprehensive evaluation.

\section*{Ethical Considerations}
The deployment of DLM-based NL2SQL systems involves critical considerations regarding data security and algorithmic bias. While our study utilizes public benchmarks, real-world applications must implement robust access control and sanitization to prevent unauthorized data exposure or malicious SQL injections. Furthermore, DLMs may inherit biases from training data, potentially leading to disparate performance across different domains or linguistic styles, which necessitates careful validation in diverse realistic database environments.

\bibliography{reference}

\clearpage
\appendix

\section{Prompt Template}
\label{sec:prompt-template}

\begin{promptbox}[System Prompt for SQL Generation]
You are a data science expert. Below, you are provided with a database schema and a natural language question. Your task is to understand the schema and generate a valid SQL query to answer the question.

Database Engine:
{db_engine}

Database Schema:
{schema}
This schema describes the database's structure, including tables, columns, primary keys, foreign keys, and any relevant relationships or constraints.

Question:
{question}

Instructions:
- Make sure you only output the information that is asked in the question. If the question asks for a specific column, make sure to only include that column in the SELECT clause, nothing more.
- The generated query should return all of the information asked in the question without any missing or extra information.
- Before generating the final SQL query, please think through the steps of how to write the query.
- Note that while the reasoning process and SQL query need to be enclosed within <answer> </answer> tag, this should not affect the quality of the SQL generation.
- The answer must contain the SQL query within ```sql ``` tags.

Output Format:
<answer>
-- Your reasoning process here
```sql
-- Your SQL query
```
</answer>

Take a deep breath and think step by step to find the correct SQL query.
\end{promptbox}

\section{Additional Experiment Settings}
\label{sec:appendix-detailed_evaluation_settings}

In this section, we provide detailed descriptions of the evaluation settings for the DLMs. Section~\ref{sec:appendix-detailed_benchmark_settings} provides the details of evaluation benchmarks. Section~\ref{sec:appendix-dlm-details} summarizes the key configuration parameters of the evaluated DLMs. Section~\ref{sec:appendix-llm-baselines} provides the settings of the LLM baselines. Section~\ref{sec:appendix-implementation_details} provides the implementation details of the evaluation settings.

\subsection{Detailed Benchmark Settings}
\label{sec:appendix-detailed_benchmark_settings}

Our evaluation spans six prominent NL2SQL benchmarks to provide a multi-dimensional assessment of DLM capabilities. \textbf{Spider} serves as the primary cross-domain benchmark, comprising 10,181 questions and 5,693 unique SQL queries across 200 databases; we evaluate on its Dev and Test sets. \textbf{BIRD} introduces large-scale, real-world challenges with 12,751 NL2SQL pairs across 95 databases from 37 domains, emphasizing database-grounded reasoning and execution efficiency. To further investigate model robustness, we incorporate three specialized variants of Spider: \textbf{Spider-DK} (535 questions, 10 databases) evaluates the integration of domain-specific knowledge, \textbf{Spider-Syn} (1,034 questions, 20 databases) tests resilience against linguistic variations through synonym substitution, and \textbf{Spider-Realistic} (508 questions, 20 databases) provides a more challenging setting by removing explicit schema mentions from natural language questions. Together, these benchmarks characterize the performance of DLMs across varying levels of structural complexity, linguistic diversity, and domain-specific requirements.

Regarding the licensing and usage of these benchmarks, we strictly adhere to the protocols defined by the original authors. The \textbf{Spider} dataset and its robustness variants, including \textbf{Spider-DK}, \textbf{Spider-Syn}, and \textbf{Spider-Realistic}, are distributed under the Creative Commons Attribution-ShareAlike 4.0 International (CC BY-SA 4.0) license. The \textbf{BIRD} benchmark is released under the Creative Commons Attribution-NonCommercial-ShareAlike 4.0 International (CC BY-NC-SA 4.0) license. All datasets are utilized exclusively for academic research purposes with appropriate attribution to the respective creators.

\subsection{Description of DLMs}
\label{sec:appendix-dlm-details}

\input{Tables/dlm_details.tex}

The core architectural configurations and inference frameworks for the evaluated DLMs are consolidated in Table~\ref{tab:dlm_details}. To further elucidate the technical foundations of these models, we provide a detailed characterization of the structural nuances and generation dynamics for each model family as follows:

\paragraph{LLaDA Series.} 
We organize the evaluated LLaDA models into two product lines that share masked denoising diffusion and parallel token refinement at inference, but correspond to different model generations. The first line comprises \textbf{LLaDA-8B}~\citep{llada} and \textbf{LLaDA-1.5-8B}~\citep{llada15}, both at 8B scale, which form the earlier LLaDA family used in our study. The second line comprises \textbf{LLaDA2.0-mini}~\citep{llada20} and \textbf{LLaDA2.1-mini}~\citep{llada21}, which constitute the LLaDA 2.x series. These models adopt a Mixture-of-Experts (MoE) architecture with 16B total parameters and approximately 1.4B active parameters per token, representing a significant update in scaling and efficiency relative to the earlier 8B checkpoints. Across both lines, decoding proceeds through iterative denoising over the full sequence rather than strict left-to-right autoregression, which supports global consistency and is well suited to structured generation tasks such as NL2SQL. The earlier LLaDA checkpoints are supported by the \texttt{dLLM} framework, whereas the LLaDA 2.x series is deployed using the \texttt{dinfer} and \texttt{dFactory} engines.

\paragraph{Dream and DreamCoder Series.} 
Dream-7B~\citep{dream} and DreamCoder-7B~\citep{dreamcoder} employ a discrete mask diffusion approach. These models treat text generation as a refinement process where tokens are progressively unmasked and denoised. This paradigm is well-suited for long-range dependency modeling and iterative structural improvement in SQL queries. Both models in this series are integrated into the \texttt{dLLM} inference framework to support efficient iterative refinement.

\paragraph{WeDLM Series.} 
The WeDLM family (7B and 8B)~\citep{wedlm} utilizes a discrete diffusion framework with a specialized hybrid decoding strategy. These models combine iterative denoising with a reranking mechanism to ensure that the final output is both semantically accurate and syntactically valid. WeDLM serves as the primary backbone for our \texttt{SQL-D1} agentic framework. The inference for the WeDLM series is powered by the dedicated \texttt{WeDLM} framework.

\paragraph{DiffuCoder.} 
DiffuCoder-7B represents the block diffusion paradigm, where text is refined in contiguous blocks rather than individual tokens. This approach aims to capture local structural constraints more effectively, which is beneficial for the rigid syntax of database queries. DiffuCoder-7B is executed within the \texttt{dLLM} framework, leveraging its support for block-based sequence refinement.

\subsection{Description of AR Baselines}
\label{sec:appendix-llm-baselines}

To provide a comprehensive performance benchmark for DLMs, we include a diverse set of AR baselines from previous works~\citep{omnisql,sqlr1}. These include leading proprietary models such as \textbf{GPT-4-Turbo} and \textbf{GPT-4o}~\citep{openaigpt4}, as well as representative open-source models across different scales and specializations, including \textbf{Qwen2.5} series (7B and 14B)~\citep{qwen2.5}, \textbf{Qwen2.5-Coder}~\citep{qwen2.5}, \textbf{DeepSeek-Coder-6.7B} (DSC)~\citep{guo2024deepseek}, \textbf{Qwen3-8B-Thinking}~\citep{yang2025qwen3} to characterize the performance gap between current DLMs and the most advanced sequential generation paradigms in the NL2SQL domain. All AR baselines are evaluated with the same retrieval-augmented context and, where applicable, the same verification and selection budgets as the corresponding DLM runs, as specified in Table~\ref{tab:main_results}.

\subsection{Implementation Details} 
\label{sec:appendix-implementation_details}

\input{Tables/ablation_full_enhancement_eval.tex}

In our experimental setup, we calibrate the inference parameters for each DLM family by building upon the official configurations provided by the respective authors. To enhance the robustness of structured reasoning, these baseline settings are further refined to accommodate the specific demands of the NL2SQL task. A uniform sampling temperature of 0.8 is applied to all models to enable multi-path exploratory reasoning, thereby facilitating the generation of a diverse candidate pool.

Majority voting and pass@$k$ are computed from execution denotations. Non-executable candidates are excluded from voting. If multiple denotation groups share the highest frequency, we select the shortest SQL as a deterministic tie-breaker. Database retrieval is restricted to the current database; for each schema field we inject exactly one offline representative example value together with schema metadata. The same retrieval setting is used for DLM and AR models in the main comparison.

Regarding the training configuration for the WeDLM series, we perform supervised fine-tuning~(SFT) on the BIRD training set. The model is trained for 5 epochs with a peak learning rate of $3.0 \times 10^{-6}$, utilizing a cosine decay schedule and a 5\% warmup phase. Architectural hyperparameters include a block size of 32. For the LLaDA2 series, our training protocol adheres to the official configurations established within the \texttt{dFactory} framework~\citep{dfactory}, which includes block-diffusion fine-tuning and specialized Mixture-of-Experts (MoE) optimization strategies. All experiments are conducted using \texttt{bf16} mixed precision to ensure numerical stability.

To facilitate inference, each NL2SQL data sample is transformed into a structured pair of input and output sequences. The input sequence integrates natural language questions with their corresponding database schemas represented in the Data Definition Language~(DDL) format. Following established practices~\citep{chess, omnisql, yang2024synthesizing, rajkumar2022evaluating}, this DDL schema is augmented with supplementary annotations, including column attribute descriptions and representative database values. Representative value annotations are included at inference.

\section{Additional Results and Analysis}
\label{sec:appendix-additional_results_and_analysis}

In this section, we provide additional results and analyses of DLM-based NL2SQL systems. Section~\ref{sec:appendix-matched_scaling} compares DLM and AR models under budget-matched test-time scaling. Section~\ref{sec:appendix-ar_sqld1} evaluates \texttt{SQL-D1} on an AR backbone. Section~\ref{sec:appendix-database_content_retrieval_impact} examines the impact of database content retrieval. Section~\ref{sec:appendix-efficiency_vs_baselines} compares inference efficiency and execution accuracy with AR baselines. Section~\ref{sec:appendix-diffusion_rendering_params} analyzes the impact of diffusion rendering parameters.

\subsection{Budget-Matched Test-time Scaling}
\label{sec:appendix-matched_scaling}

Table~\ref{tab:matched_scaling} compares LLaDA2.1-mini and Qwen2.5-Coder-7B under the same prompt, retrieval context, candidate budget, and execution-based voting rule, with $k{=}1$ as greedy decoding. Under this matched setting, LLaDA2.1-mini shows a steeper Maj@$k$ gain than Qwen2.5-Coder-7B ($+8.8$ / $+9.9$ versus $+5.6$ / $+7.3$ at $k{=}4$ / $8$). The same trend appears in Table~\ref{tab:main_results}: on Spider-Test, LLaDA2.1-mini improves from 71.4 to 78.6~($+7.2$), while Qwen2.5-Coder-7B improves from 82.2 to 85.6~($+3.4$). Relative to greedy decoding, LLaDA2.1-mini also obtains a larger coverage lift to pass@8 ($48.8$ to $67.1$, $+18.3$) than Qwen2.5-Coder-7B ($50.9$ to $66.1$, $+15.2$). WeDLM-8B behaves differently: its Maj@8 gain is small ($+1.6$), while the pass@8$-$Maj@8 gap is the largest among the three models ($10.6$ versus $7.9$ for Qwen2.5-Coder-7B and $8.3$ for LLaDA2.1-mini), suggesting remaining headroom for better selection under high-variance diffusion generation.

\input{Tables/matched_scaling_eval.tex}

\subsection{\texttt{SQL-D1} on an AR Backbone}
\label{sec:appendix-ar_sqld1}

Table~\ref{tab:ar_sqld1_eval} reports a matched \texttt{SQL-D1} ablation on BIRD-Dev for WeDLM-7B, WeDLM-8B, and Qwen2.5-7B under the same module definitions. The greedy baseline in this table excludes retrieval, matching the first row of Table~\ref{tab:ablation_enhancement_eval}; greedy scores in Table~\ref{tab:main_results} include retrieval and correspond to the Retriever row. \texttt{SQL-D1} improves both backbones: the full-pipeline gain is $+8.8$ for WeDLM-8B and $+8.7$ for Qwen2.5-7B, and WeDLM-8B attains a slightly higher full-pipeline EX~(54.9 vs.\ 53.2). The larger single-pass variance of diffusion generation continues to benefit from retrieval, verification, and selection, as further reflected by the scaling gaps in Appendix~\ref{sec:appendix-matched_scaling}.

\input{Tables/ar_sqld1_eval.tex}

\subsection{Impact of Database Content Retrieval}
\label{sec:appendix-database_content_retrieval_impact}

\input{Tables/ablation_efficiency.tex}

\begin{figure*}[t!]
  \centering
  \includegraphics[width=\linewidth]{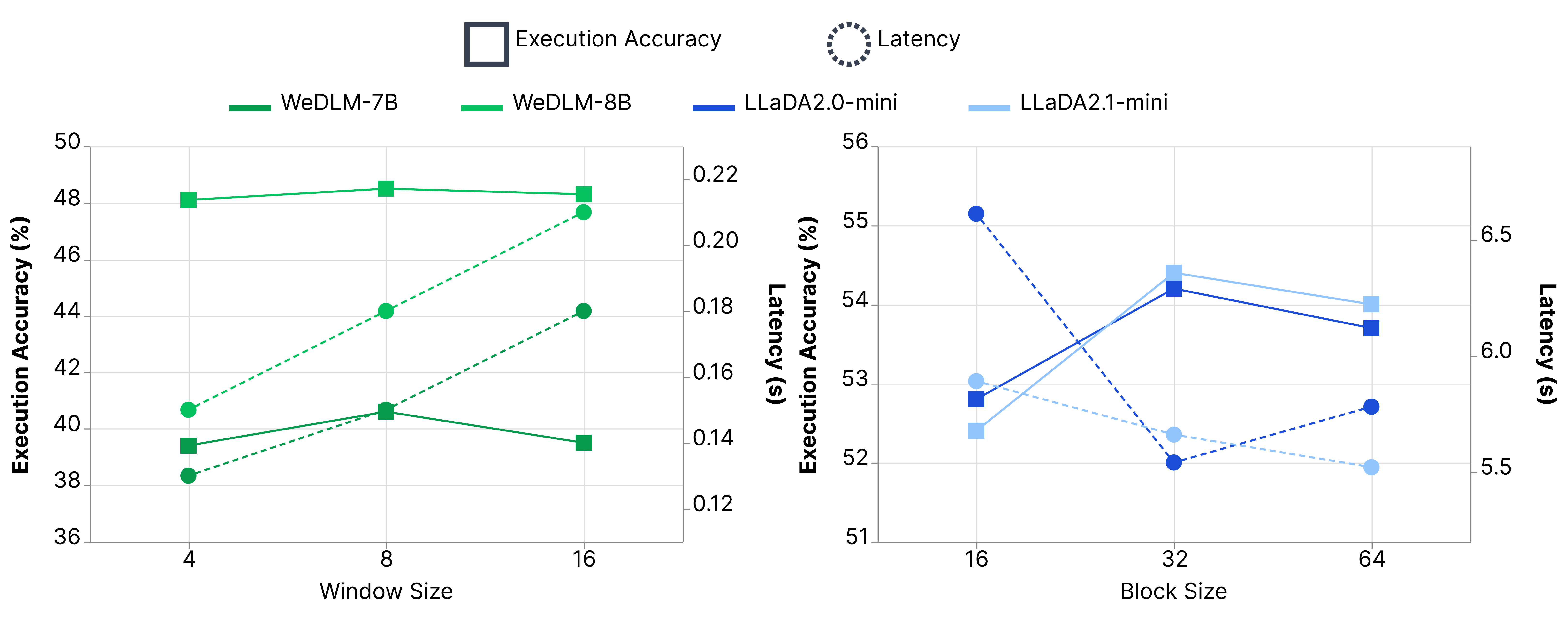}
  \caption{Impact of diffusion rendering hyperparameters on execution accuracy and inference latency on BIRD-Dev~(\textbf{left}: window size for WeDLM; \textbf{right}: block size for LLaDA2). Solid lines with square markers denote execution accuracy~(left axis); dashed lines with circle markers denote latency~(right axis).}
  \label{fig:diffusion_rendering_params}
\end{figure*}

Table~\ref{tab:ablation_full_enhancement_eval} provides a comprehensive ablation analysis investigating the influence of database content retrieval across diverse DLM architectures and benchmarks. The empirical evidence demonstrates that the integration of database content retrieval consistently enhances performance across all evaluated models, highlighting the importance of grounded contextual information for diffusion-based NL2SQL reasoning. Notably, the LLaDA2 series shows substantial improvements, with LLaDA2.1-mini exhibiting an execution accuracy gain of 3.8\% on Spider-Test and 4.3\% on BIRD-Dev. Furthermore, the Dream and DiffuCoder series exhibit a robust response to the inclusion of database literals, particularly on more complex benchmarks such as Spider-Realistic. These observations suggest that explicit database content effectively alleviates the schema linking difficulties inherent in the non-autoregressive generation process, thereby facilitating a more precise alignment between natural language queries and the underlying database distributions.

\input{Tables/error_taxonomy.tex}

\subsection{Efficiency vs. AR Baselines}
\label{sec:appendix-efficiency_vs_baselines}

Table~\ref{tab:ablation_efficiency} provides a comparative analysis of inference efficiency, computational cost, and execution accuracy between DLMs and representative autoregressive baselines on the BIRD-Dev dataset. The results demonstrate that DLMs have achieved significant potential in the NL2SQL task, with the WeDLM series exhibiting exceptional efficiency profiles. Specifically, WeDLM-8B achieves a latency of 0.18 seconds, substantially lower than the 0.28 seconds of Qwen2.5-7B, while using a comparable token budget (2.2K vs.\ 2.1K) and maintaining a competitive execution accuracy of 53.0\%. Furthermore, LLaDA2.1-mini reaches the best EX among the evaluated DLMs (58.7\%), surpassing the reasoning-enhanced Qwen3-8B-Thinking (51.8\%) and performing on par with Qwen2.5-Coder-7B (58.2\%). These findings highlight that the WeDLM architecture, by leveraging inference infrastructure partially compatible with traditional LLMs, successfully bridges the gap between high-speed non-autoregressive generation and the precision required for complex database reasoning. The versatile trade-offs offered by these diffusion architectures provide a compelling alternative to sequential generation paradigms in real-time application scenarios.

\subsection{Impact of Diffusion Rendering Parameters}
\label{sec:appendix-diffusion_rendering_params}

Figure~\ref{fig:diffusion_rendering_params} investigates how inference-time diffusion rendering hyperparameters shape the coupling between execution accuracy and inference efficiency on BIRD-Dev. Rendering governs the amount of context exposed and processed at each denoising step: an overly narrow rendering budget may leave relevant schema and question context under-specified, whereas an excessively wide budget expands the rendered span and materially increases per-query computation. Across the two strongest DLM families in our study, this tension manifests in distinct but consistent patterns.

For the WeDLM series~(\textbf{left}), we sweep the context window size from 4 to 16. Execution accuracy is comparatively insensitive to this range for WeDLM-8B, which remains between 48.1\% and 48.5\%, yet WeDLM-7B improves from 39.4\% at window size 4 to a peak of 40.6\% at window size 8 before reverting to 39.5\% at window size 16, indicating that a moderately expanded window is beneficial whereas further expansion yields no additional gain. By contrast, inference latency increases monotonically with window size for both models, from 0.13\,s to 0.18\,s on WeDLM-7B and from 0.15\,s to 0.21\,s on WeDLM-8B, confirming that richer rendering primarily purchases latency rather than accuracy once the window is sufficiently large. For the LLaDA2 series~(\textbf{right}), block size exerts a stronger influence on accuracy. Both LLaDA2.0-mini and LLaDA2.1-mini attain their highest execution accuracy at block size 32 (54.2\% and 54.4\%, respectively), improving over the smaller block setting at 16 (52.8\% and 52.4\%) while exhibiting slight degradation at block size 64 (53.7\% and 54.0\%). Latency responses differ across variants: LLaDA2.0-mini is slowest at block size 16 (6.61\,s) and fastest at block size 32 (5.54\,s), whereas LLaDA2.1-mini shows a steady reduction in latency as block size grows, from 5.89\,s at 16 to 5.52\,s at 64. Together, these curves suggest that insufficient rendering can suppress accuracy, while additional rendering beyond an architecture-dependent sweet spot chiefly inflates inference time with diminishing returns on execution accuracy.

\section{Error Case Study}
\label{sec:appendix-case_study}

Building upon the fine-grained failure-mode analysis in Section~\ref{sec:experiments} and the distributional summary in Figure~\ref{fig:error_distribution}, this section supplements aggregate error statistics with interpretable, question-level post-mortems of DLM-generated SQL. 
The exposition adheres to the established five-category taxonomy detailed in Table~\ref{tab:error_taxonomy}, which provides a structured classification of failure modes spanning schema grounding, semantic alignment, and structural integrity.
This framework foregrounds prototypical errors through a side-by-side presentation of model outputs, succinct diagnostic commentary, and corrected formulations when such contrasts clarify the intended query semantics.

\subsection{Table Error}

\begin{casebox}[Case 1: Table Mismatch Error]

\textbf{Question.} How many negative comments did Neil McGuigan get in his posts?

\textbf{Predicted SQL.}
\begin{alltt}
\textcolor{blue}{SELECT} COUNT(*) 
\textcolor{blue}{FROM} comments c 
\textcolor{blue}{JOIN} users u \textcolor{blue}{ON} c.UserId = u.Id 
\textcolor{blue}{WHERE} c.Score < 60 
\textcolor{blue}{AND} u.DisplayName = 'Neil McGuigan'
\end{alltt}

\vspace{0.5em}
\textbf{Issue.} \textit{The query counts comments made by Neil McGuigan, but the NL query asks for negative comments on his posts. The join should be on PostId to posts, not on UserId to users. This is a Table Mismatch error because the wrong table is joined (users instead of posts). This is a \textbf{Table} error.}

\vspace{0.5em}

\textbf{Corrected SQL.}
\begin{alltt}
\textcolor{blue}{SELECT} COUNT(T3.Id) 
\textcolor{blue}{FROM} users \textcolor{blue}{AS} T1 
\textcolor{blue}{INNER JOIN} posts \textcolor{blue}{AS} T2 
        \textcolor{blue}{ON} T1.Id = T2.OwnerUserId 
\textcolor{blue}{INNER JOIN} comments \textcolor{blue}{AS} T3 
        \textcolor{blue}{ON} T2.Id = T3.PostId 
\textcolor{blue}{WHERE} T1.DisplayName = 'Neil McGuigan' 
       \textcolor{blue}{AND} T3.Score < 60
\end{alltt}

\end{casebox}

\subsection{Value Error}

\begin{casebox}[Case 2: Value Mismatch Error]

\textbf{Question.} Give the names of the schools with the percent eligible for free meals in K-12 is more than 0.1 and test takers whose test score is greater than or equal to 1500?

\textbf{Predicted SQL.}
\begin{alltt}
\textcolor{blue}{SELECT} s.sname 
\textcolor{blue}{FROM} frpm f \textcolor{blue}{JOIN} satscores s \textcolor{blue}{ON} 
     f.CDSCode = s.cds
\textcolor{blue}{WHERE} f.
        `Percent (%) Eligible Free (K-12)`
      > 0.1 
\textcolor{blue}{AND} \textcolor{red}{s.NumGE1500} \textcolor{blue}{>=} \textcolor{red}{1500} \scalebox{0.8}{\textcolor{red}{\XSolidBrush}}
\end{alltt}

\vspace{0.5em}
\textbf{Issue.} \textit{The query uses `s.NumGE1500 >= 1500` to filter schools with test scores >= 1500, but `NumGE1500` is the count of students scoring >= 1500, not the test score itself. The correct condition should involve the average scores or a derived condition. This is a value mismatch error because the value used in the condition is incorrect.}
\vspace{0.5em}

\textbf{Corrected SQL.}
\begin{alltt}
\textcolor{blue}{SELECT} T2.`School Name` 
\textcolor{blue}{FROM} satscores \textcolor{blue}{AS} T1 
\textcolor{blue}{INNER JOIN} frpm \textcolor{blue}{AS} T2 \textcolor{blue}{ON} 
           T1.cds = T2.CDSCode 
\textcolor{blue}{WHERE} \textcolor{blue}{CAST}(T2.`Free Meal Count (K-12)` 
      \textcolor{blue}{AS} REAL) 
         / T2.`Enrollment (K-12)` 
         > 0.1 
\textcolor{blue}{AND} T1.NumGE1500 > 0
\end{alltt}
\end{casebox}

\subsection{Condition Error}

\begin{casebox}[Case 3: Attribute Mismatch Error]
\textbf{Question.} Which active district has the highest average score in Reading?

\textbf{Predicted SQL.}
\begin{alltt}
\textcolor{blue}{SELECT} \textcolor{red}{s.District} \scalebox{0.8}{\textcolor{red}{\XSolidBrush}}
\textcolor{blue}{FROM} \textcolor{red}{satscores s} 
\textcolor{blue}{JOIN} schools sch \textcolor{blue}{ON} s.cds = sch.CDSCode
\textcolor{blue}{WHERE} sch.StatusType = 'Active'
\textcolor{blue}{GROUP BY} \textcolor{red}{s.District} \scalebox{0.8}{\textcolor{red}{\XSolidBrush}}
\textcolor{blue}{ORDER BY} \textcolor{blue}{AVG}(s.AvgScrRead) \textcolor{blue}{DESC}
\textcolor{blue}{LIMIT} 1
\end{alltt}

\vspace{0.5em}
\textbf{Issue.} \textit{The query uses s.District in the \texttt{SELECT} and \texttt{GROUP BY} clauses, but the satscores table does not have a 'District' column; it has 'dname' for district name. This is an attribute mismatch error, which is a \textbf{Condition} error.}
\vspace{0.5em}

\textbf{Corrected SQL.}
\begin{alltt}
\textcolor{blue}{SELECT} \textcolor{green}{T1.District} \textcolor{green}{\checkmark}
\textcolor{blue}{FROM} \textcolor{green}{schools AS T1} \textcolor{green}{\checkmark}
\textcolor{blue}{INNER JOIN} satscores \textcolor{blue}{AS} T2 \textcolor{blue}{ON} 
           T1.CDSCode = T2.cds 
\textcolor{blue}{WHERE} T1.StatusType = 'Active' 
\textcolor{blue}{ORDER BY} T2.AvgScrRead \textcolor{blue}{DESC} 
\textcolor{blue}{LIMIT} 1
\end{alltt}
\end{casebox}

\begin{casebox}[Case 4: Explicit Condition Missing]
\textbf{Question.} In which country was the first European Grand Prix hosted? Name the circuit and location.

\textbf{Predicted SQL.}
\begin{alltt}
\textcolor{blue}{SELECT} c.country, c.name, c.location 
\textcolor{blue}{FROM} races r \textcolor{blue}{JOIN} circuits c 
     \textcolor{blue}{ON} r.circuitId = c.circuitId
\textcolor{blue}{WHERE} r.year = (
      \textcolor{blue}{SELECT} MIN(year) 
      \textcolor{blue}{FROM} races)
\end{alltt}

\vspace{0.5em}
\textbf{Issue.} \textit{The query correctly identifies the first European Grand Prix by using MIN(year) but fails to filter for races named 'European Grand Prix'. The condition r.year = (SELECT MIN(year) FROM races) selects the earliest race overall, not specifically the first European Grand Prix. This is an Explicit Condition Missing error because the required condition to filter by race name is missing.}
\vspace{0.5em}

\textbf{Corrected SQL.}
\begin{alltt}
\textcolor{blue}{SELECT} T1.country, T1.location 
\textcolor{blue}{FROM} circuits \textcolor{blue}{AS} T1 
\textcolor{blue}{INNER JOIN} races \textcolor{blue}{AS} T2 
       \textcolor{blue}{ON} T2.circuitID = T1.circuitId 
\textcolor{blue}{WHERE} \textcolor{green}{T2.name = 'European Grand Prix'} \textcolor{green}{\checkmark}
\textcolor{blue}{ORDER BY} T2.year \textcolor{blue}{ASC} 
\textcolor{blue}{LIMIT} 1
\end{alltt}
\end{casebox}

\subsection{Function Error}

\begin{casebox}[Case 5: Function Error]
\textbf{Question.} What is the average fastest lap time of Lewis Hamilton?

\textbf{Predicted SQL.}
\begin{alltt}
\textcolor{blue}{SELECT} AVG(
    \textcolor{red}{STRFTIME('%s', r.fastestLapTime)}) 
    \textcolor{blue}{AS} average_fastest_lap_time_seconds
\textcolor{blue}{FROM} results r 
\textcolor{blue}{JOIN} drivers d 
    \textcolor{blue}{ON} r.driverId = d.driverId
\textcolor{blue}{WHERE} d.forename = 'Lewis' 
    \textcolor{blue}{AND} d.surname = 'Hamilton'
\end{alltt}

\vspace{0.5em}
\textbf{Issue.} \textit{The SQL query uses STRFTIME('\%s', r.fastestLapTime) to convert the fastest lap time to seconds, but the evidence states the time is recorded in 'MM:SS.mmm' format. STRFTIME('\%s') expects a time in 'HH:MM:SS' format and will return NULL or incorrect results for 'MM:SS.mmm'. The correct approach would be to parse the minutes and seconds separately, e.g., using SUBSTR and CAST. This is a Date/Time Function error.}
\vspace{0.5em}

\textbf{Corrected SQL.}
\begin{alltt}
\textcolor{blue}{SELECT} AVG(
  \textcolor{green}{CAST}(\textcolor{green}{SUBSTR}( \textcolor{green}{\checkmark}
    T2.fastestLapTime, 1, 
    \textcolor{blue}{INSTR}(T2.fastestLapTime, ':') - 1) 
    \textcolor{blue}{AS} INTEGER) * 60 + 
  \textcolor{green}{CAST}(\textcolor{green}{SUBSTR}( \textcolor{green}{\checkmark}
    T2.fastestLapTime, 
    \textcolor{blue}{INSTR}(T2.fastestLapTime, ':') + 1) 
    \textcolor{blue}{AS} REAL)) 
\textcolor{blue}{FROM} drivers \textcolor{blue}{AS} T1 
\textcolor{blue}{INNER JOIN} results \textcolor{blue}{AS} T2 
\textcolor{blue}{ON} T1.driverId = T2.driverId 
\textcolor{blue}{WHERE} T1.surname = 'Hamilton' 
  \textcolor{blue}{AND} T1.forename = 'Lewis'
\end{alltt}
\end{casebox}

\end{document}

%% file: Tables/main_eval.tex
\begin{table*}[t!]
	\centering
	\footnotesize
	\setlength{\tabcolsep}{0pt}
	\setlength{\heavyrulewidth}{0.08em}
	\setlength{\lightrulewidth}{0.03em}
	\setlength{\cmidrulewidth}{0.03em}
	\renewcommand{\arraystretch}{0.95}
	\begin{tabular*}{\textwidth}{@{\extracolsep{\fill}} l cccccccccccc}
	\toprule
	\multirow{2}{*}{\textbf{Models}} &
	  \multicolumn{2}{c}{\shortstack{\textbf{Spider}\\[-2pt]\textbf{(Dev)}}} &
	  \multicolumn{2}{c}{\shortstack{\textbf{Spider}\\[-2pt]\textbf{(Test)}}} &
	  \multicolumn{2}{c}{\shortstack{\textbf{BIRD}\\[-2pt]\textbf{(Dev)}}} &
	  \multicolumn{2}{c}{\shortstack{\textbf{Spider-}\\[-2pt]\textbf{DK}}} &
	  \multicolumn{2}{c}{\shortstack{\textbf{Spider-}\\[-2pt]\textbf{Syn}}} &
	  \multicolumn{2}{c}{\shortstack{\textbf{Spider-}\\[-2pt]\textbf{Realistic}}} \\
	\cmidrule(lr){2-3}\cmidrule(lr){4-5}\cmidrule(lr){6-7}\cmidrule(lr){8-9}\cmidrule(lr){10-11}\cmidrule(lr){12-13}
	&
	  \textbf{Gre} &
	  \textbf{Maj} &
	  \textbf{Gre} &
	  \textbf{Maj} &
	  \textbf{Gre} &
	  \textbf{Maj} &
	  \textbf{Gre} &
	  \textbf{Maj} &
	  \textbf{Gre} &
	  \textbf{Maj} &
	  \textbf{Gre} &
	  \textbf{Maj} \\
	\midrule
	\multicolumn{13}{l}{\textit{Diffusion Language Models}} \\
	\midrule
	Dream-7B           & 49.8 & 54.2 & 51.0 & 53.2 & 13.1 & 19.4 & 34.0 & 45.4 & 48.9 & 53.9 & 46.0 & 51.2 \\
	DreamCoder-7B      & 47.0 & 54.2 & 47.2 & 52.5 & 25.7 & 29.9 & 27.9 & 36.1 & 31.1 & 43.7 & 28.9 & 39.2 \\
	DiffuCoder-7B 	   & 48.9 & 66.0 & 51.3 & 63.7 & 35.4 & 37.8 & 41.2 & 56.8 & 49.8 & 65.4 & 47.6 & 63.2 \\
	WeDLM-7B           & 60.9 & 70.3 & 67.0 & 73.7 & 39.7 & 43.6 & 45.5 & 56.3 & 60.9 & 66.4 & 51.2 & 59.3 \\
	WeDLM-8B           & 66.6 & 73.1 & 68.0 & 72.9 & \textbf{51.4} & 53.0 & \textbf{54.0} & 61.7 & \textbf{69.3} & 72.9 & \underline{63.0} & 66.1 \\
	LLaDA-8B           & 50.2 & 60.6 & 52.6 & 61.9 & 24.0 & 33.6 & 43.2 & 55.3 & 51.2 & 61.4 & 48.0 & 58.9 \\
	 LLaDA-1.5-8B        & 51.1 & 61.1 & 51.9 & 62.7 & 26.6 & 33.8 & 42.8 & 52.0 & 59.9 & 62.1 & 49.0 & 58.1 \\
	LLaDA2.0-mini 	   & \underline{68.2} & \textbf{74.8} & \underline{71.1} & \underline{78.1} & 48.2 & \underline{55.5} & \underline{53.3} & \textbf{67.7} & \underline{69.2} & \underline{75.4} & 62.0 & \textbf{74.6} \\
	LLaDA2.1-mini 	   & \textbf{68.5} & \underline{74.6} & \textbf{71.4} & \textbf{78.6} & \underline{48.8} & \textbf{58.7} & 52.5 & \underline{67.5} & 68.7 & \textbf{76.3} & \textbf{63.8} & \underline{72.2} \\
	\midrule
	\multicolumn{13}{l}{\textit{Autoregressive  Language Models}} \\
	\midrule
	GPT-4-Turbo        & 72.4 & 72.2 & \underline{83.4} & 84.2 & \textbf{62.0} & 63.6 & 72.3 & 72.1 & 62.9 & 63.5 & \underline{67.5} & 68.3 \\
	GPT-4o             & 70.9 & 70.7 & 83.2 & 84.9 & \underline{61.9} & \underline{64.0} & \underline{72.9} & 73.5 & 59.6 & 62.3 & 66.5 & 66.7 \\
	DeepSeek-Coder-6.7B           & 63.2 & 63.2 & 70.5 & 73.2 & 43.1 & 48.0 & 60.9 & 64.1 & 49.9 & 51.7 & 58.7 & 58.9 \\
	Qwen2.5-7B         & 65.4 & 68.9 & 76.8 & 82.6 & 46.9 & 56.4 & 63.7 & 71.8 & 54.2 & 60.0 & 56.7 & 63.6 \\
	Qwen2.5-Coder-7B   & \underline{73.4} & \underline{77.1} & 82.2 & \underline{85.6} & 50.9 & 58.2 & 67.5 & 73.6 & \underline{63.1} & \underline{66.9} & 66.7 & \underline{70.5} \\
	Qwen3-8B-Thinking  &   -   &   -   &   -   &   -   & 49.1 & 51.8 &   -   &   -   &   -   &   -   &   -   &   -   \\
	Qwen2.5-14B        & 66.5 & 69.7 & 82.0 & 84.0 & 56.7 & 62.1 & 72.3 & \underline{74.0} & 58.1 & 60.7 & 62.4 & 65.2 \\
	Qwen2.5-Coder-14B  & \textbf{78.1} & \textbf{80.6} & \textbf{86.6} & \textbf{88.0} & 61.5 & \textbf{66.1} & \textbf{73.6} & \textbf{77.8} & \textbf{68.2} & \textbf{69.3} & \textbf{76.2} & \textbf{74.2} \\
	\bottomrule
	\end{tabular*}

	\caption{Performance comparison of DLMs and AR models on main NL2SQL benchmarks. \textbf{Gre} refers to greedy search, \textbf{Maj} refers to majority voting by 8 candidates. Overall inference settings are kept the same for all models. Within DLMs and AR models, the best and second-best results are in \textbf{bold} and \underline{underlined}, respectively.}
	\label{tab:main_results}
\end{table*}

%% file: Tables/difficulty_eval.tex
\begin{table}[t!]
	\centering
	\small
	\setlength{\tabcolsep}{0pt}
	\begin{tabular*}{\linewidth}{@{\extracolsep{\fill}} lcccc}
	\toprule
	\textbf{Models} & \textbf{Sim.} & \textbf{Mod.} & \textbf{Chall.} & \textbf{All} \\
	\midrule
	\multicolumn{5}{l}{\textit{Diffusion Language Models}} \\
	Dream-7B      & 27.9 & 7.1 & 4.1 & 19.4 \\
	DreamCoder-7B & 38.5 & 17.9 & 13.1 & 29.9  \\
	DiffuCoder-7B & 46.5 & 24.4 & 24.8 & 37.8  \\
	LLaDA-8B      & 44.7 & 17.2 & 15.9 & 33.6  \\
	 LLaDA-1.5-8B   & 44.7 & 17.0 & 17.9 & 33.8  \\
	WeDLM-7B      & 52.5 & 30.6 & 27.6 & 43.6  \\
	WeDLM-8B      & 60.8 & 42.2 & 37.9 & 53.0  \\
	LLaDA2.0-mini & 62.4 & 45.7 & 42.8 & 55.5  \\
	LLaDA2.1-mini & 64.5 & 50.2 & 49.0 & 58.7  \\
	\midrule
	\multicolumn{5}{l}{\textit{Specific NL2SQL Methods}} \\
	CodeS-15B     & 65.8 & 48.8 & 42.4 & 58.5 \\
	DAIL-SQL      & 63.0 & 45.6 & 43.1 & 55.9 \\
	SuperSQL      & 66.9 & 46.5 & 43.8 & 58.5 \\
	\bottomrule
	\end{tabular*}
	\caption{Execution accuracy (\%) of different complexity levels on BIRD-Dev dataset. \textbf{Sim.} refers to Simple, \textbf{Mod.} refers to Moderate, \textbf{Chall.} refers to Challenging.}
	\label{tab:difficulty_eval}
\end{table}

%% file: Tables/ablation_sqld1_eval.tex
\begin{table*}[ht!]
	\centering
	\begin{tabular}{lcccccc}
	\toprule
	 & \shortstack{\textbf{Spider}\\\textbf{(Dev)}} & \shortstack{\textbf{Spider}\\\textbf{(Test)}} & \shortstack{\textbf{BIRD}\\\textbf{(Dev)}} & \shortstack{\textbf{Spider-}\\\textbf{DK}} & \shortstack{\textbf{Spider-}\\\textbf{Syn}} & \shortstack{\textbf{Spider-}\\\textbf{Realistic}} \\
	\midrule
	WeDLM-7B                        			& 58.6 & 56.7 & 38.4 & 42.5 & 59.6 & 50.1 \\
	- w/ Retriever $A_r$ / DB Retrieval			& 60.9 & 67.0 & 39.7 & 45.5 & 60.9 & 51.2 \\
	- w/ Verifier~($A_r$ + $A_g$ + $A_v$)		& 70.5 & 73.9 & 44.1 & - & - & - \\
	- w/ Selector~($A_r$ + $A_g$ + $A_s$)		& 70.7 & 74.3 & 44.6& - & - & - \\
	- w/ \texttt{SQL-D1}~($A_r$ + $A_g$ + $A_v$ + $A_s$) & \cellcolor{cyan!12}\textbf{70.8} & \cellcolor{cyan!12}\textbf{74.6} & \cellcolor{cyan!12}\textbf{45.0} & \cellcolor{cyan!12}\textbf{62.0} & \cellcolor{cyan!12}\textbf{70.6} & \cellcolor{cyan!12}\textbf{64.8} \\
	\midrule
	WeDLM-8B                        			& 63.1 & 66.3 & 46.1 & 53.0 & 66.7 & 59.9 \\
	- w/ Retriever $A_r$ / DB Retrieval			& 66.6 & 68.0 & 51.4 & 54.0 & 69.3 & 63.0 \\
	- w/ Verifier~($A_r$ + $A_g$ + $A_v$)		& 73.9 & 74.6 & 53.8 & - & - & - \\
	- w/ Selector~($A_r$ + $A_g$ + $A_s$)		& 74.4 & 74.9 & 54.5 & - & - & - \\
	- w/ \texttt{SQL-D1}~($A_r$ + $A_g$ + $A_v$ + $A_s$) & \cellcolor{cyan!12}\textbf{75.0} & \cellcolor{cyan!12}\textbf{75.2} & \cellcolor{cyan!12}\textbf{54.9} & \cellcolor{cyan!12}\textbf{67.8} & \cellcolor{cyan!12}\textbf{74.2} & \cellcolor{cyan!12}\textbf{70.5} \\
	\bottomrule
	\end{tabular}
	\caption{Ablation study of \texttt{SQL-D1} on main benchmarks. Baseline results are reported by greedy search generation.}
	\label{tab:ablation_enhancement_eval}
\end{table*}

%% file: Tables/ablation_training_eval.tex
\begin{table}[t]
	\centering
	\scalebox{0.90}{%
	\begin{tabular}{lc}
	\toprule
	\textbf{Models} & \textbf{EX~(\%)~/~$\Delta$~(\%)} \\
	\midrule
	WeDLM-7B-Instruct  & 39.7 \\
	- w/ SFT~(\texttt{BIRD-Train-9K})  & 40.2~($\uparrow$~0.5) \\
	\cmidrule{2-2}
	WeDLM-8B-Instruct  & 51.4 \\
	- w/ SFT~(\texttt{BIRD-Train-9K})  & 49.3~($\downarrow$~2.1) \\
	\cmidrule{2-2}
	LLaDA2.0-mini  & 48.2 \\
	- w/ SFT~(\texttt{BIRD-Train-9K})  & 46.4~($\downarrow$~1.8) \\
	\midrule
	\end{tabular}%
	}
	\caption{SFT ablation of WeDLM and LLaDA2.0-mini on BIRD-Dev. Overall results are reported with greedy search generation.}
	\label{tab:ablation_training_eval}
\end{table}

%% file: Tables/dlm_details.tex
\begin{table*}[t]
    \centering
    \scriptsize
    \setlength{\tabcolsep}{3.2pt}
    \resizebox{0.92\textwidth}{!}{%
    \begin{tabular}{lcccc}
    \toprule
    \textbf{Model} &
    \textbf{Params} &
    \textbf{Diffusion Paradigm} &
    \textbf{Inference Framework} \\
    \midrule
    Dream-7B      & 7B   & Discrete mask diffusion    & \texttt{dLLM} \\
    DreamCoder-7B & 7B   & Discrete mask diffusion    & \texttt{dLLM} \\
    DiffuCoder-7B & 7B   & Block diffusion            & \texttt{dLLM} \\
    LLaDA-8B      & 8B   & Masked denoising diffusion & \texttt{dLLM} \\
     LLaDA-1.5-8B  & 8B   & Masked denoising diffusion & \texttt{dLLM} \\
    LLaDA2.0-mini & 16B (MoE)   & Masked denoising diffusion & \texttt{dinfer}/\texttt{dFactory} \\
    LLaDA2.1-mini & 16B (MoE)   & Masked denoising diffusion & \texttt{dinfer}/\texttt{dFactory} \\
    WeDLM-7B      & 7B   & Discrete diffusion         & \texttt{WeDLM} \\
    WeDLM-8B      & 8B   & Discrete diffusion         & \texttt{WeDLM} \\
    \bottomrule
    \end{tabular}%
    }
    \caption{Diffusion Language Models used in this work and their key configuration parameters.}
    \label{tab:dlm_details}
\end{table*}

%% file: Tables/ablation_full_enhancement_eval.tex
\begin{table*}[t!]
	\centering
	\begin{tabular}{lcccccc}
	\toprule
	 & \shortstack{\textbf{Spider}\\\textbf{(Dev)}} & \shortstack{\textbf{Spider}\\\textbf{(Test)}} & \shortstack{\textbf{BIRD}\\\textbf{(Dev)}} & \shortstack{\textbf{Spider-}\\\textbf{DK}} & \shortstack{\textbf{Spider-}\\\textbf{Syn}} & \shortstack{\textbf{Spider-}\\\textbf{Realistic}} \\
	\midrule
	Dream-7B                        & 52.5 & 52.1 & 18.8 & 42.4 & 52.4 & 48.0 \\
	- w/ DB Retrieval 				& 54.2 & 53.2 & 19.4 & 45.4 & 53.9 & 51.2 \\
	\cmidrule(l){2-7}
	DreamCoder-7B                   & 52.1 & 50.2 & 27.9 & 31.6 & 38.7 & 29.3 \\
	- w/ DB Retrieval 				& 54.2 & 52.5 & 29.9 & 36.1 & 43.7 & 39.2 \\
	\cmidrule(l){2-7}
	DiffuCoder-7B 					& 62.3 & 62.6 & 36.2 & 54.6 & 63.2 & 54.5 \\	
	- w/ DB Retrieval 				& 66.0 & 63.7 & 37.8 & 56.8 & 65.4 & 63.2 \\
	\cmidrule(l){2-7}
	LLaDA-8B                        & 60.1 & 59.9 & 33.2 & 53.6 & 60.1 & 53.0 \\
	- w/ DB Retrieval 				& 60.6 & 61.9 & 33.6 & 55.3 & 61.4 & 58.9 \\
	\cmidrule(l){2-7}
	LLaDA-1.5-8B                    & 60.8 & 60.3 & 33.3 & 48.7 & 61.1 & 52.4 \\
	- w/ DB Retrieval 				& 61.1 & 62.7 & 33.8 & 52.0 & 62.1 & 58.1 \\
	\cmidrule(l){2-7}
	LLaDA2.0-mini                   & 73.0 & 75.9 & 54.2 & 60.0 & 73.1 & 65.5 \\
	- w/ DB Retrieval 				& 74.8 & 78.1 & 55.5 & 67.7 & 75.4 & 74.6 \\
	\cmidrule(l){2-7}
	LLaDA2.1-mini                   & 72.0 & 74.8 & 54.4 & 61.1 & 74.0 & 69.1 \\
	- w/ DB Retrieval 				& 74.6 & 78.6 & 58.7 & 67.5 & 76.3 & 72.2 \\	
	\bottomrule
	\end{tabular}
	\caption{Ablation study of database content retrieval on main NL2SQL benchmarks.}
	\label{tab:ablation_full_enhancement_eval}
\end{table*}

%% file: Tables/matched_scaling_eval.tex
\begin{table*}[t]
	\centering
	\small
	\setlength{\tabcolsep}{10pt}
	\begin{tabular}{lcccccc}
	\toprule
	\textbf{Model} &
	  \textbf{Gre} &
	  \textbf{Maj@4} &
	  \textbf{Maj@8} &
	  \textbf{$\Delta$ Maj@8} &
	  \textbf{pass@8} &
	  \textbf{pass@8$-$Maj@8} \\
	\midrule
	LLaDA2.1-mini      & 48.8 & 57.6 & 58.7 & +9.9 & 67.1 & 8.3 \\
	Qwen2.5-Coder-7B   & 50.9 & 56.5 & 58.2 & +7.3 & 66.1 & 7.9 \\
	WeDLM-8B           & 51.4 & --   & 53.0 & +1.6 & 63.6 & 10.6 \\
	\bottomrule
	\end{tabular}
	\caption{Budget-matched test-time scaling on BIRD-Dev~(EX\,\%). All models use the same prompt, retrieval context, and execution-based majority voting. $k{=}1$ corresponds to greedy decoding. $\Delta$ Maj@8 is the gain relative to greedy. Maj@4 is not reported for WeDLM-8B.}
	\label{tab:matched_scaling}
\end{table*}

%% file: Tables/ar_sqld1_eval.tex
\begin{table*}[t]
	\centering
	\small
	\setlength{\tabcolsep}{10pt}
	\begin{tabular}{lccc}
	\toprule
	\textbf{Setting} & \textbf{WeDLM-7B} & \textbf{WeDLM-8B} & \textbf{Qwen2.5-7B} \\
	\midrule
	Greedy baseline & 38.4 & 46.1 & 44.5 \\
	+ Retriever $A_r$ / DB Retrieval & 39.7 & 51.4 & 46.9 \\
	+ Verifier~($A_r$ + $A_g$ + $A_v$) & 44.1 & 53.8 & 51.5 \\
	+ Selector~($A_r$ + $A_g$ + $A_s$) & 44.6 & 54.5 & 52.4 \\
	+ \texttt{SQL-D1}~($A_r$ + $A_g$ + $A_v$ + $A_s$) & \textbf{45.0} & \textbf{54.9} & \textbf{53.2} \\
	\midrule
	Delta~(full \texttt{SQL-D1} minus Greedy) & +6.6 & +8.8 & +8.7 \\
	\bottomrule
	\end{tabular}
	\caption{\texttt{SQL-D1} ablation on BIRD-Dev~(EX\,\%) under matched module definitions for DLM and AR backbones. The greedy baseline excludes retrieval, consistent with the first row of Table~\ref{tab:ablation_enhancement_eval}. Greedy scores in Table~\ref{tab:main_results} include retrieval and correspond to the Retriever row. Delta is full \texttt{SQL-D1} minus the greedy baseline.}
	\label{tab:ar_sqld1_eval}
\end{table*}

%% file: Tables/ablation_efficiency.tex
\begin{table}[t!]
	\centering
	\resizebox{0.95\linewidth}{!}{%
	\begin{tabular}{lccc}
	\toprule
	\textbf{Models} & 
	\textbf{Latency (s)} & 
	\textbf{Token~(K)} &
	\textbf{EX~(\%)} \\
	\midrule
	\multicolumn{4}{l}{\textit{Autoregressive Language Models}} \\
	Qwen2.5-7B & 0.28 & 2.1 & 56.4 \\
	Qwen2.5-Coder-7B & 0.31 & 2.5 & 58.2 \\
	Qwen3-8B-Thinking & 1.02 & 5.6 & 51.8 \\
	\midrule
	\multicolumn{4}{l}{\textit{Diffusion Language Models}} \\
	WeDLM-7B & 0.14 & 1.7 & 43.6 \\
	WeDLM-8B & 0.18 & 2.2 & 53.0 \\
	LLaDA2.0-mini & 5.54 & 3.6 & 55.5 \\
	LLaDA2.1-mini & 5.66 & 3.9 & 58.7 \\
	\bottomrule
	\end{tabular}%
	}
	\caption{Efficiency and execution accuracy trade-offs of DLMs on BIRD-Dev datasets.}
	\label{tab:ablation_efficiency}
\end{table}

%% file: Tables/error_taxonomy.tex
\begin{table*}[t!]
	\centering
	\renewcommand{\arraystretch}{1.2}
	\begin{tabular}{llp{10cm}}
	\toprule
	\textbf{Category} & \textbf{Error Type} & \textbf{Description} \\
	\midrule
	\multirow{2}{*}{\textbf{Table}} & Table Mismatch & Selection of incorrect tables from the database schema. \\
	& Table Missing & Omission of essential tables required for the query. \\
	\midrule
	\textbf{Value} & Value Mismatch & Incorrect extraction, formatting, or alignment of database literals (e.g., strings, dates) with the natural language evidence. \\
	\midrule
	\multirow{2}{*}{\textbf{Condition}} & Attribute Error & Selection of incorrect columns within \texttt{WHERE} or \texttt{HAVING} clauses. \\
	& Operator Error & Use of incorrect logical (e.g., \texttt{AND} vs. \texttt{OR}) or comparison operators (e.g., \texttt{>} vs. \texttt{<}). \\
	\midrule
	\textbf{Function} & Aggregation Error & Incorrect application, omission, or selection of aggregate functions such as \texttt{SUM}, \texttt{AVG}, or \texttt{COUNT}. \\
	\midrule
	\multirow{2}{*}{\textbf{Others}} & Clause Missing & Absence of critical structural components such as \texttt{GROUP BY}, \texttt{ORDER BY}, or \texttt{LIMIT}. \\
	& Structural Error & Incorrect formulation of complex nested queries and subqueries. \\
	\bottomrule
	\end{tabular}
	\caption{Taxonomy of errors exhibited by DLMs in NL2SQL tasks, categorized into five primary failure modes.}
	\label{tab:error_taxonomy}
\end{table*}

%% file: reference.bib
@article{dailsql,
author = {Dawei Gao and
Haibin Wang and
Yaliang Li and
Xiuyu Sun and
Yichen Qian and
Bolin Ding and
Jingren Zhou},
title = {Text-to-SQL Empowered by Large Language Models: A Benchmark Evaluation},
journal = {CoRR},
volume = {abs/2308.15363},
year = {2023}
}

@article{DBLP:journals/corr/abs-2408-05109,
  author       = {Xinyu Liu and
                  Shuyu Shen and
                  Boyan Li and
                  Peixian Ma and
                  Runzhi Jiang and
                  Yuyu Luo and
                  Yuxin Zhang and
                  Ju Fan and
                  Guoliang Li and
                  Nan Tang},
  title        = {A Survey of {NL2SQL} with Large Language Models: Where are we, and
                  where are we going?},
  journal      = {CoRR},
  volume       = {abs/2408.05109},
  year         = {2024}
}

@article{c3sql,
  title={C3: Zero-shot text-to-sql with chatgpt},
  author={Dong, Xuemei and Zhang, Chao and Ge, Yuhang and Mao, Yuren and Gao, Yunjun and Lin, Jinshu and Lou, Dongfang and others},
  journal={arXiv preprint arXiv:2307.07306},
  year={2023}
}

@article{dtssql,
  title={DTS-SQL: Decomposed Text-to-SQL with Small Large Language Models},
  author={Pourreza, Mohammadreza and Rafiei, Davood},
  journal={arXiv preprint arXiv:2402.01117},
  year={2024}
}

@inproceedings{resdsql,
author = {Haoyang Li and Jing Zhang and Cuiping Li and Hong Chen},
title = "RESDSQL: Decoupling Schema Linking and Skeleton Parsing for Text-to-SQL",
booktitle = "AAAI",
year = "2023"
}

@article{openaigpt4,
title={GPT-4 Technical Report},
author={OpenAI},
journal={arXiv preprint arXiv:2303.08774},
year={2023}
}

@misc{bird,
title={Can LLM Already Serve as A Database Interface? A BIg Bench for Large-Scale Database Grounded Text-to-SQLs},
author={Jinyang Li and Binyuan Hui and Ge Qu and Binhua Li and Jiaxi Yang and Bowen Li and Bailin Wang and Bowen Qin and Ruiying Geng and Nan Huo and Xuanhe Zhou and Chenhao Ma and Guoliang Li and Kevin C. C. Chang and Fei Huang and Reynold Cheng and Yongbin Li},
year={2023},
eprint={2305.03111},
archivePrefix={arXiv},
primaryClass={cs.CL}
}

@inproceedings{spider,
title = {Spider: A Large-Scale Human-Labeled Dataset for Complex and Cross-Domain Semantic Parsing and Text-to-SQL Task},
author = {Tao Yu and Rui Zhang and Kai Yang and Michihiro Yasunaga and Dongxu Wang and Zifan Li and James Ma and Irene Li and Qingning Yao and Shanelle Roman and Zilin Zhang and Dragomir Radev},
booktitle = "Proceedings of the 2018 Conference on Empirical Methods in Natural Language Processing",
address = "Brussels, Belgium",
publisher = "Association for Computational Linguistics",
year = 2018
}

@misc{macsql,
title={MAC-SQL: A Multi-Agent Collaborative Framework for Text-to-SQL},
author={Bing Wang and Changyu Ren and Jian Yang and Xinnian Liang and Jiaqi Bai and Linzheng Chai and Zhao Yan and Qian-Wen Zhang and Di Yin and Xing Sun and Zhoujun Li},
year={2024},
eprint={2312.11242},
archivePrefix={arXiv},
primaryClass={cs.CL}
}

@article{chess,
  title={Chess: Contextual harnessing for efficient sql synthesis},
  author={Talaei, Shayan and Pourreza, Mohammadreza and Chang, Yu-Chen and Mirhoseini, Azalia and Saberi, Amin},
  journal={arXiv preprint arXiv:2405.16755},
  year={2024}
}

@article{codes,
  title={Codes: Towards building open-source language models for text-to-sql},
  author={Li, Haoyang and Zhang, Jing and Liu, Hanbing and Fan, Ju and Zhang, Xiaokang and Zhu, Jun and Wei, Renjie and Pan, Hongyan and Li, Cuiping and Chen, Hong},
  journal={Proceedings of the ACM on Management of Data},
  volume={2},
  number={3},
  pages={1--28},
  year={2024},
  publisher={ACM New York, NY, USA}
}

@article{ma2024plug,
  title={A Plug-and-Play Natural Language Rewriter for Natural Language to SQL},
  author={Ma, Peixian and Li, Boyan and Jiang, Runzhi and Fan, Ju and Tang, Nan and Luo, Yuyu},
  journal={arXiv preprint arXiv:2412.17068},
  year={2024}
}

@article{sqlo1,
  title={SQL-o1: A Self-Reward Heuristic Dynamic Search Method for Text-to-SQL},
  author={Lyu, Shuai and Luo, Haoran and Ou, Zhonghong and Zhu, Yifan and Shang, Xiaoran and Qin, Yang and Song, Meina},
  journal={arXiv preprint arXiv:2502.11741},
  year={2025}
}

@article{mctssql,
  title={MCTS-SQL: An Effective Framework for Text-to-SQL with Monte Carlo Tree Search},
  author={Yuan, Shuozhi and Chen, Liming and Yuan, Miaomiao and Zhao, Jin and Peng, Haoran and Guo, Wenming},
  journal={arXiv preprint arXiv:2501.16607},
  year={2025}
}

@article{omnisql,
  title={OmniSQL: Synthesizing High-quality Text-to-SQL Data at Scale},
  author={Li, Haoyang and Wu, Shang and Zhang, Xiaokang and Huang, Xinmei and Zhang, Jing and Jiang, Fuxin and Wang, Shuai and Zhang, Tieying and Chen, Jianjun and Shi, Rui and others},
  journal={arXiv preprint arXiv:2503.02240},
  year={2025}
}

@article{supersql,
  title={The dawn of natural language to SQL: are we fully ready?},
  author={Li, Boyan and Luo, Yuyu and Chai, Chengliang and Li, Guoliang and Tang, Nan},
  journal={arXiv preprint arXiv:2406.01265},
  year={2024}
}

@article{chasesql,
  title={Chase-sql: Multi-path reasoning and preference optimized candidate selection in text-to-sql},
  author={Pourreza, Mohammadreza and Li, Hailong and Sun, Ruoxi and Chung, Yeounoh and Talaei, Shayan and Kakkar, Gaurav Tarlok and Gan, Yu and Saberi, Amin and Ozcan, Fatma and Arik, Sercan O},
  journal={arXiv preprint arXiv:2410.01943},
  year={2024}
}

@article{opensearchsql,
  title={OpenSearch-SQL: Enhancing Text-to-SQL with Dynamic Few-shot and Consistency Alignment},
  author={Xie, Xiangjin and Xu, Guangwei and Zhao, Lingyan and Guo, Ruijie},
  journal={arXiv preprint arXiv:2502.14913},
  year={2025}
}

@article{yang2024synthesizing,
  title={Synthesizing text-to-SQL data from weak and strong LLMs},
  author={Yang, Jiaxi and Hui, Binyuan and Yang, Min and Yang, Jian and Lin, Junyang and Zhou, Chang},
  journal={arXiv preprint arXiv:2408.03256},
  year={2024}
}

@article{rajkumar2022evaluating,
  title={Evaluating the text-to-sql capabilities of large language models},
  author={Rajkumar, Nitarshan and Li, Raymond and Bahdanau, Dzmitry},
  journal={arXiv preprint arXiv:2204.00498},
  year={2022}
}

@article{hong2024next,
  title={Next-generation database interfaces: A survey of llm-based text-to-sql},
  author={Hong, Zijin and Yuan, Zheng and Zhang, Qinggang and Chen, Hao and Dong, Junnan and Huang, Feiran and Huang, Xiao},
  journal={arXiv preprint arXiv:2406.08426},
  year={2024}
}

@article{route,
  title={ROUTE: Robust Multitask Tuning and Collaboration for Text-to-SQL},
  author={Qin, Yang and Chen, Chao and Fu, Zhihang and Chen, Ze and Peng, Dezhong and Hu, Peng and Ye, Jieping},
  journal={arXiv preprint arXiv:2412.10138},
  year={2024}
}

@article{maamari2024death,
  title={The death of schema linking? text-to-sql in the age of well-reasoned language models},
  author={Maamari, Karime and Abubaker, Fadhil and Jaroslawicz, Daniel and Mhedhbi, Amine},
  journal={arXiv preprint arXiv:2408.07702},
  year={2024}
}

@article{qwen2.5,
  title={Qwen2. 5-coder technical report},
  author={Hui, Binyuan and Yang, Jian and Cui, Zeyu and Yang, Jiaxi and Liu, Dayiheng and Zhang, Lei and Liu, Tianyu and Zhang, Jiajun and Yu, Bowen and Lu, Keming and others},
  journal={arXiv preprint arXiv:2409.12186},
  year={2024}
}

@article{xiyansql,
  title={Xiyan-sql: A multi-generator ensemble framework for text-to-sql},
  author={Gao, Yingqi and Liu, Yifu and Li, Xiaoxia and Shi, Xiaorong and Zhu, Yin and Wang, Yiming and Li, Shiqi and Li, Wei and Hong, Yuntao and Luo, Zhiling and others},
  journal={arXiv e-prints},
  pages={arXiv--2411},
  year={2024}
}

@inproceedings{sqlr1,
 author = {Ma, Peixian and Zhuang, Xialie and Xu, Chengjin and Jiang, Xuhui and Chen, Ran and Guo, Jian},
 booktitle = {Advances in Neural Information Processing Systems},
 editor = {D. Belgrave and C. Zhang and H. Lin and R. Pascanu and P. Koniusz and M. Ghassemi and N. Chen},
 pages = {174505--174537},
 publisher = {Curran Associates, Inc.},
 title = {SQL-R1: Training Natural Language to SQL Reasoning Model By Reinforcement Learning},
 url = {https://proceedings.neurips.cc/paper_files/paper/2025/file/ff00aae6ce2b5e519ca4cf769f0c4a47-Paper-Conference.pdf},
 volume = {38},
 year = {2025}
}

@misc{spider-dk,
      title={Exploring Underexplored Limitations of Cross-Domain Text-to-SQL Generalization}, 
      author={Yujian Gan and Xinyun Chen and Matthew Purver},
      year={2021},
      eprint={2109.05157},
      archivePrefix={arXiv},
      primaryClass={cs.CL}
}

@inproceedings{spider-syn,
    title = "Towards Robustness of Text-to-{SQL} Models against Synonym Substitution",
    author = "Gan, Yujian  and
      Chen, Xinyun  and
      Huang, Qiuping  and
      Purver, Matthew  and
      Woodward, John R.  and
      Xie, Jinxia  and
      Huang, Pengsheng",
    month = aug,
    year = "2021",
    address = "Online",
    publisher = "Association for Computational Linguistics",
    url = "https://aclanthology.org/2021.acl-long.195",
    doi = "10.18653/v1/2021.acl-long.195",
    pages = "2505--2515",
}

@article{spider-realistic,
  title={Structure-grounded pretraining for text-to-sql},
  author={Deng, Xiang and Awadallah, Ahmed Hassan and Meek, Christopher and Polozov, Oleksandr and Sun, Huan and Richardson, Matthew},
  journal={arXiv preprint arXiv:2010.12773},
  year={2020}
}

@article{agenterscalesql,
  title={Agentar-Scale-SQL: Advancing Text-to-SQL through Orchestrated Test-Time Scaling},
  author={Wang, Pengfei and Sun, Baolin and Dong, Xuemei and Dai, Yaxun and Yuan, Hongwei and Chu, Mengdie and Gao, Yingqi and Qi, Xiang and Zhang, Peng and Yan, Ying},
  journal={arXiv preprint arXiv:2509.24403},
  year={2025}
}

@article{llada,
  title={Large Language Diffusion Models},
  author={Nie, Shen and Zhu, Fengqi and You, Zebin and Zhang, Xiaolu and Ou, Jingyang and Hu, Jun and Zhou, Jun and Lin, Yankai and Wen, Ji-Rong and Li, Chongxuan},
  journal={arXiv preprint arXiv:2502.09992},
  year={2025}
}

@article{llada15,
  title={LLaDA 1.5: Variance-Reduced Preference Optimization for Large Language Diffusion Models},
  author={Zhu, Fengqi and Wang, Rongzhen and Nie, Shen and Zhang, Xiaolu and Wu, Chunwei and Hu, Jun and Zhou, Jun and Chen, Jianfei and Lin, Yankai and Wen, Ji-Rong and others},
  journal={arXiv preprint arXiv:2505.19223},
  year={2025}
}

@article{wedlm,
  title={WeDLM: Reconciling Diffusion Language Models with Standard Causal Attention for Fast Inference},
  author={Liu, Aiwei and He, Minghua and Zeng, Shaoxun and Zhang, Linhao and Wu, Chuhan and Jia, Wei and Liu, Yuan and Yu, Yang and Zhou, Xiao and Zhou, Jie},
  journal={arXiv preprint arXiv:2512.22737},
  year={2025}
}

@article{dream,
  title={Dream 7B: Diffusion Large Language Models},
  author={Ye, Jiacheng and Xie, Zhihui and Zheng, Lin and Gao, Jiahui and Wu, Zirui and Jiang, Xin and Li, Zhenguo and Kong, Lingpeng},
  journal={arXiv preprint arXiv:2508.15487},
  year={2025}
}

@article{dreamcoder,
  title={Dream-coder 7b: An open diffusion language model for code},
  author={Xie, Zhihui and Ye, Jiacheng and Zheng, Lin and Gao, Jiahui and Dong, Jingwei and Wu, Zirui and Zhao, Xueliang and Gong, Shansan and Jiang, Xin and Li, Zhenguo and others},
  journal={arXiv preprint arXiv:2509.01142},
  year={2025}
}

@misc{llada21,
      title={LLaDA2.1: Speeding Up Text Diffusion via Token Editing}, 
      author={Tiwei Bie and Maosong Cao and Xiang Cao and Bingsen Chen and Fuyuan Chen and Kun Chen and Lun Du and Daozhuo Feng and Haibo Feng and Mingliang Gong and Zhuocheng Gong and Yanmei Gu and Jian Guan and Kaiyuan Guan and Hongliang He and Zenan Huang and Juyong Jiang and Zhonghui Jiang and Zhenzhong Lan and Chengxi Li and Jianguo Li and Zehuan Li and Huabin Liu and Lin Liu and Guoshan Lu and Yuan Lu and Yuxin Ma and Xingyu Mou and Zhenxuan Pan and Kaida Qiu and Yuji Ren and Jianfeng Tan and Yiding Tian and Zian Wang and Lanning Wei and Tao Wu and Yipeng Xing and Wentao Ye and Liangyu Zha and Tianze Zhang and Xiaolu Zhang and Junbo Zhao and Da Zheng and Hao Zhong and Wanli Zhong and Jun Zhou and Junlin Zhou and Liwang Zhu and Muzhi Zhu and Yihong Zhuang},
      year={2026},
      eprint={2602.08676},
      archivePrefix={arXiv},
      primaryClass={cs.LG},
      url={https://arxiv.org/abs/2602.08676}, 
}

@misc{llada20,
      title={LLaDA2.0: Scaling Up Diffusion Language Models to 100B}, 
      author={Tiwei Bie and Maosong Cao and Kun Chen and Lun Du and Mingliang Gong and Zhuochen Gong and Yanmei Gu and Jiaqi Hu and Zenan Huang and Zhenzhong Lan and Chengxi Li and Chongxuan Li and Jianguo Li and Zehuan Li and Huabin Liu and Ling Liu and Guoshan Lu and Xiaocheng Lu and Yuxin Ma and Jianfeng Tan and Lanning Wei and Ji-Rong Wen and Yipeng Xing and Xiaolu Zhang and Junbo Zhao and Da Zheng and Jun Zhou and Junlin Zhou and Zhanchao Zhou and Liwang Zhu and Yihong Zhuang},
      year={2025},
      eprint={2512.15745},
      archivePrefix={arXiv},
      primaryClass={cs.LG},
      url={https://arxiv.org/abs/2512.15745}, 
}

@article{dlmsurvey,
  title={A Survey on Diffusion Language Models},
  author={Li, Tianyi and Chen, Mingda and Guo, Bowei and Shen, Zhiqiang},
  journal={arXiv preprint arXiv:2508.10875},
  year={2025}
}

@inproceedings{arriola2025interpolating,
  title={Interpolating Autoregressive and Discrete Denoising Diffusion Language Models},
  author={Marianne Arriola and Aaron Gokaslan and Justin T Chiu and Jiaqi Han and Zhihan Yang and Zhixuan Qi and Subham Sekhar Sahoo and Volodymyr Kuleshov},
  booktitle={The Thirteenth International Conference on Learning Representations},
  year={2025},
  url={https://openreview.net/forum?id=tyEyYT267x}
}

@article{yu2025discrete,
  title={Discrete Diffusion in Large Language and Multimodal Models: A Survey},
  author={Yu, Runpeng and Li, Qi and Wang, Xinchao},
  journal={arXiv preprint arXiv:2506.13759},
  year={2025}
}

@misc{zhou2026dllm,
      title={dLLM: Simple Diffusion Language Modeling}, 
      author={Zhanhui Zhou and Lingjie Chen and Hanghang Tong and Dawn Song},
      year={2026},
      eprint={2602.22661},
      archivePrefix={arXiv},
      primaryClass={cs.CL},
      url={https://arxiv.org/abs/2602.22661}, 
}

@article{dinfer,
    title={dInfer: An Efficient Inference Framework for Diffusion Language Models},
    author={Yuxin Ma and Lun Du and Lanning Wei and Kun Chen and Qian Xu and Kangyu Wang and Guofeng Feng and Guoshan Lu and Lin Liu and Xiaojing Qi and Xinyuan Zhang and Zhen Tao and Haibo Feng and Ziyun Jiang and Ying Xu and Zenan Huang and Yihong Zhuang and Haokai Xu and Jiaqi Hu and Zhenzhong Lan and Junbo Zhao and Jianguo Li and Da Zheng},
    year={2025},
    journal={arXiv preprint arXiv:2510.08666}
}

@misc{zhang2026expertchoiceroutingenablesadaptive,
      title={Expert-Choice Routing Enables Adaptive Computation in Diffusion Language Models}, 
      author={Shuibai Zhang and Caspian Zhuang and Chihan Cui and Zhihan Yang and Fred Zhangzhi Peng and Yanxin Zhang and Haoyue Bai and Zack Jia and Yang Zhou and Guanhua Chen and Ming Liu},
      year={2026},
      eprint={2604.01622},
      archivePrefix={arXiv},
      primaryClass={cs.LG},
      url={https://arxiv.org/abs/2604.01622}, 
}

@misc{mounier2025reviewremaskrefiner3,
      title={Review, Remask, Refine (R3): Process-Guided Block Diffusion for Text Generation}, 
      author={Nikita Mounier and Parsa Idehpour},
      year={2025},
      eprint={2507.08018},
      archivePrefix={arXiv},
      primaryClass={cs.CL},
      url={https://arxiv.org/abs/2507.08018}, 
}

@misc{vonrütte2025generalizedinterpolatingdiscretediffusion,
      title={Generalized Interpolating Discrete Diffusion}, 
      author={Dimitri von Rütte and Janis Fluri and Yuhui Ding and Antonio Orvieto and Bernhard Schölkopf and Thomas Hofmann},
      year={2025},
      eprint={2503.04482},
      archivePrefix={arXiv},
      primaryClass={cs.CL},
      url={https://arxiv.org/abs/2503.04482}, 
}

@inproceedings{gong2025scaling,
    title={Scaling Diffusion Language Models via Adaptation from Autoregressive Models},
    author={Shansan Gong and Shivam Agarwal and Yizhe Zhang and Jiacheng Ye and Lin Zheng and Mukai Li and Chenxin An and Peilin Zhao and Wei Bi and Jiawei Han and Hao Peng and Lingpeng Kong},
    booktitle={The Thirteenth International Conference on Learning Representations},
    year={2025},
    url={https://openreview.net/forum?id=j1tSLYKwg8}
}

@article{yang2025qwen3,
  title={Qwen3 technical report},
  author={Yang, An and Li, Anfeng and Yang, Baosong and Zhang, Beichen and Hui, Binyuan and Zheng, Bo and Yu, Bowen and Gao, Chang and Huang, Chengen and Lv, Chenxu and others},
  journal={arXiv preprint arXiv:2505.09388},
  year={2025}
}

@article{guo2024deepseek,
  title={DeepSeek-Coder: when the large language model meets programming--the rise of code intelligence},
  author={Guo, Daya and Zhu, Qihao and Yang, Dejian and Xie, Zhenda and Dong, Kai and Zhang, Wentao and Chen, Guanting and Bi, Xiao and Wu, Yifan and Li, YK and others},
  journal={arXiv preprint arXiv:2401.14196},
  year={2024}
}

@misc{nie2025scalingmaskeddiffusionmodels,
      title={Scaling up Masked Diffusion Models on Text}, 
      author={Shen Nie and Fengqi Zhu and Chao Du and Tianyu Pang and Qian Liu and Guangtao Zeng and Min Lin and Chongxuan Li},
      year={2025},
      eprint={2410.18514},
      archivePrefix={arXiv},
      primaryClass={cs.AI},
      url={https://arxiv.org/abs/2410.18514}, 
}

@misc{ni2025trainingoptimallargediffusion,
      title={Training Optimal Large Diffusion Language Models}, 
      author={Jinjie Ni and Qian Liu and Chao Du and Longxu Dou and Hang Yan and Zili Wang and Tianyu Pang and Michael Qizhe Shieh},
      year={2025},
      eprint={2510.03280},
      archivePrefix={arXiv},
      primaryClass={cs.LG},
      url={https://arxiv.org/abs/2510.03280}, 
}

@misc{dfactory,
  title={dFactory: Easy and Efficient dLLM Fine-Tuning},
  author={inclusionAI},
  year={2026},
  url={https://github.com/inclusionAI/dFactory}
}

@article{zhang2025corrective,
  title={Corrective Diffusion Language Models},
  author={Zhang, Shuibai and Peng, Fred Zhangzhi and Zhang, Yiheng and Pan, Jin and Chrysos, Grigorios G},
  journal={arXiv preprint arXiv:2512.15596},
  year={2025}
}

@inproceedings{leafsql,
  author    = {Zhao Tan and
               Xiping Liu and
               Qing Shu and
               Qizhi Wan and
               Dexi Liu and
               Changxuan Wan},
  title     = {LEAF-SQL: Level-wise Exploration with Adaptive Fine-graining for Text-to-SQL Skeleton Prediction},
  booktitle = {Proceedings of the 42nd IEEE International Conference on Data Engineering (ICDE)},
  year      = {2026}
}

@article{Soni_2026,
   title={Schema on the Inside: A Two-Phase Fine-Tuning Method for High-Efficiency Text-to-SQL at Scale},
   volume={40},
   ISSN={2159-5399},
   url={http://dx.doi.org/10.1609/aaai.v40i47.41446},
   DOI={10.1609/aaai.v40i47.41446},
   number={47},
   journal={Proceedings of the AAAI Conference on Artificial Intelligence},
   publisher={Association for the Advancement of Artificial Intelligence (AAAI)},
   author={Soni, Chinmay and Chourasia, Shivam and Kumar, Gaurav and Kapoor, Hitesh},
   year={2026},
   month=Mar, pages={40110–40117} }

@misc{marssql,
      title={MARS-SQL: A multi-agent reinforcement learning framework for Text-to-SQL}, 
      author={Haolin Yang and Jipeng Zhang and Zhitao He and Alexander Zhou and Yi R. Fung},
      year={2026},
      eprint={2511.01008},
      archivePrefix={arXiv},
      primaryClass={cs.CL},
      url={https://arxiv.org/abs/2511.01008}, 
}

@article{diver,
   title={DIVER: A Robust Text-to-SQL System with Dynamic Interactive Value Linking and Evidence Reasoning},
   volume={4},
   ISSN={2836-6573},
   url={http://dx.doi.org/10.1145/3786640},
   DOI={10.1145/3786640},
   number={1},
   journal={Proceedings of the ACM on Management of Data},
   publisher={Association for Computing Machinery (ACM)},
   author={Nan, Yafeng and Sun, Haifeng and Zhuang, Zirui and Qi, Qi and Chu, Guojun and Liao, Jianxin and Pei, Dan and Wang, Jingyu},
   year={2026},
   month=Apr, pages={1–24} }

@misc{memosql,
      title={Memo-SQL: Structured Decomposition and Experience-Driven Self-Correction for Training-Free NL2SQL}, 
      author={Zerui Yang and Weichuan Wang and Yanwei Xu and Linqi Song and Yudai Matsuda and Wei Han and Bo Bai},
      year={2026},
      eprint={2601.10011},
      archivePrefix={arXiv},
      primaryClass={cs.AI},
      url={https://arxiv.org/abs/2601.10011}, 
}

@article{cai2026text2sqlflow,
  title={TEXT2SQL-FLOW: A Robust SQL-Aware Data Augmentation Framework for Text-to-SQL},
  author={Cai, Qifeng and Liang, Hao and Xu, Chang and Xie, Tao and Zhang, Wentao and Cui, Bin},
  journal={arXiv preprint arXiv:2511.10192v4},
  year={2026}
}

@misc{yang2026continuousdiffusionscalescompetitively,
      title={Continuous Diffusion Scales Competitively with Discrete Diffusion for Language}, 
      author={Zhihan Yang and Wei Guo and Shuibai Zhang and Subham Sekhar Sahoo and Yongxin Chen and Arash Vahdat and Morteza Mardani and John Thickstun},
      year={2026},
      eprint={2605.18530},
      archivePrefix={arXiv},
      primaryClass={cs.CL},
      url={https://arxiv.org/abs/2605.18530}, 
}
